\documentclass[twocolumn,prb]{revtex4}
\usepackage{amsfonts}

\usepackage{graphicx}
\usepackage{epsfig}
\usepackage{amsmath, amssymb}
\usepackage{verbatim}
\usepackage{bm}
\usepackage{multirow}

\def\sgn{{\rm sgn}}
\def\Ref{Ref. \onlinecite}

\begin{document}

\title{Genons, twist defects, and projective non-Abelian braiding statistics}
\author{Maissam Barkeshli}
\author{Chao-Ming Jian}
\author{Xiao-Liang Qi}
\affiliation{Department of Physics, Stanford University, Stanford, CA 94305 }

\begin{abstract}
It has recently been realized that a general class of non-abelian defects can be created in conventional
topological states by introducing extrinsic defects, such as lattice dislocations or superconductor-ferromagnet
domain walls in conventional quantum Hall states or topological insulators. In this paper,
we begin by placing these defects within the broader conceptual scheme of extrinsic twist
defects associated with symmetries of the topological state. We explicitly study several classes of examples,
including $Z_2$ and $Z_3$ twist defects, where the topological state with $N$ twist defects
can be mapped to a topological state without twist defects on a genus $g \propto N$ surface. To
emphasize this connection we refer to the twist defects as \it genons\rm. We develop methods to compute the
projective non-abelian braiding statistics of the genons, and
%, both from a one-dimensional Luttinger liquid perspective, and from a bulk geometric perspective.
we find the braiding is given by adiabatic modular transformations, or Dehn twists, of the topological
state on the effective genus $g$ surface. We study the relation between this projective braiding statistics and the ordinary
non-abelian braiding statistics obtained when the genons become deconfined, finite-energy excitations.
We find that the braiding is generally different, in contrast to the Majorana case, which opens the possibility for fundamentally novel behavior.
%as the projective statistics is subject to fewer constraints than ordinary non-abelian statistics.
%As an example, we
We find situations where the genons have quantum dimension $2$ and can be used for universal topological quantum computing
(TQC), while the host topological state is by itself non-universal for TQC. %, contradicting a conjecture for ordinary non-abelian statistics.
\end{abstract}
\maketitle

\section{Introduction}

Some of the most important discoveries in condensed matter physics over the last few decades have been about topological states of matter.\cite{wen04,nayak2008,qiRMP2011, hasan2010} Topological states form a class of quantum many-body states that are distinguished by principles of topology rather than symmetry, which implies that they have physical properties that are robust under arbitrary local perturbations and that are not associated with any local order parameter. Well known examples of such topological properties include the existence of gapless robust edge states, topology-dependent ground state degeneracies, fractional charge and statistics of topological quasi-particles, and quantized response properties. For some topological states, known as non-Abelian states,\cite{nayak2008}  there are a finite number of degenerate lowest energy states for any given configuration of topological quasi-particles. The degeneracy between these states is topologically protected, being robust against any local perturbation. An adiabatic motion of the quasi-particles in real space therefore leads to a nontrivial unitary operation in the space of degenerate ground states, which is described by the non-Abelian statistics of the quasiparticles. Such a state has been proposed to be used, if realized, for intrinsically fault-tolerant, robust quantum information storage and processing: the quantum information is stored in the topologically degenerate states, and the operations are realized by motion of quasiparticles. This idea, known as topological quantum computation (TQC),\cite{nayak2008} has been a major driving force for the study of topological states of matter.

The discussion of topological quasiparticles in topological states can be extended to \it extrinsic \rm defects, which are point-like objects that
%Rather than search for true non-abelian quasiparticles, it has been realized that \it projective \rm non-abelian statistics may actually be much easier to realize and manipulate. Projective non-abelian statistics arises when the non-abelian defects
are not intrinsic, finite-energy excitations of the system, but instead have a long-ranged confining interaction with each other.\cite{read2000,alicea2012review,bombin2010,kitaev2012,barkeshli2012prx,you2012,you2012b,teo2010,freedman2011,swingle2011} Such defects can be created and controlled by an external field, and they may also carry non-abelian statistics, similar to intrinsic topological quasiparticles. However the overall phase of the unitary operation generated by braiding the defects is generically non-universal and path dependent, due to their long ranged interaction. Therefore the statistics of the extrinsic defects is only well-defined up to a phase, implying that their braiding forms a \it projective \rm representation of the braid group. This possibility is referred to as {\it projective} non-Abelian statistics.\footnote{Projective representations of the permutation group
were proposed for arbitrary dimensions in \Ref{wilczek1998}, and later shown to be inconsistent with locality in \Ref{read2003}. A different possibility, called projective ribbon permutation statistics, was found to occur 3+1 dimensions.\cite{teo2010,freedman2011} In this paper, we instead concentrate on projective representations of the braid group, which has received little attention in the physics literature. }
A simple example of a non-trivial extrinsic defect is the vortex in a two-dimensional $p_x+ip_y$ topological superfluid\cite{read2000}, which has Majorana zero modes and non-Abelian statistics similar to the topological quasi-particles in the Moore-Read Pfaffian fractional quantum Hall (FQH) state.\cite{Moore1991} The overall phase of the statistics is undetermined due to the logarithmic interaction of the vortices. Recently, similar extrinsic defects with Majorana zero modes have been discussed in several other physical systems\cite{alicea2012review}.

%such as vortices in a thin-film superconductor, or extrinsic defects introduced into the system. Currently, the most intense focus is on the simplest kind of projective non-abelian statistics: the Majorana fermions in topological superconductors,\cite{alicea2012review} which are intimately related to the non-abelian statistics of quasiparticles in the Moore-Read Pfaffian FQH state.\cite{Moore1991, read2000}
Since projective non-abelian statistics are not described by the same mathematical framework\cite{preskillLectures} as intrinsic topological quasiparticles,
there may potentially be a world of possibilities that would be inconsistent for ordinary non-abelian statistics. An important direction therefore is to develop
the theory of projective non-abelian braiding statistics and to investigate the novel possibilities.

A second important direction is to explore ways of obtaining exotic non-Abelian defects, beyond Majorana fermions, in simple, experimentally achievable settings.
This is motivated in part by the well-known deficiency of Majorana fermions to yield a universal gate set for TQC.\cite{preskillLectures}
Recently, extrinsic defects with projective non-Abelian statistics beyond Majorana fermions have been proposed in several systems.
In fractional Chern insulators ({\it i.e.} FQH states in lattice models without an external magnetic field)\cite{sun2011,neupert2011,tang2011,sheng2011, regnault2011,thomale2012}, a class of states called topological nematic states can be realized when the Chern number of the partially occupied energy band is larger than $1$.\cite{barkeshli2012prx} Topological nematic states are topologically equivalent to conventional multi-layer FQH states on regular lattices, but lattice dislocations effectively change the topology and introduce ``worm holes" between the two layers, which leads to the non-Abelian statistics of the dislocations. Similar non-Abelian lattice dislocations can also be realized in $Z_N$ rotor models.\cite{bombin2010,you2012, kitaev2012}. %XLnote: Pls check the references. Shall we also cite kitaev and Liang Kong also?
A different type of extrinsic defect has been proposed on the boundary of FQH or fractional quantum spin Hall (FQSH) states, at domain walls between regions in which the edge states obtain different mass terms.\cite{lindner2012,clarke2012,cheng2012, vaezi2012}

%As will be shown in this work, these two types of defects proposed recently are closely related.
%
%a family of such non-Abelian defects have been discovered in several Abelian topological states, including in abelian topological states
%Recently, more exotic examples of projective non-abelian
%statistics, beyond Majorana fermions, have been proposed in , such as
%fractional Chern insulators ({\it i.e.} FQH states in lattice models without external magnetic field)\cite{barkeshli2011}, conventional FQH and fractional quantum spin Hall (FQSH) states\cite{lindner2012,clarke2012,cheng2012, vaezi2012}, and $Z_N$ rotor
%models.\cite{bombin2010,kitaev2012,you2012}
%In particular, Ref. \cite{barkeshli2011} proposed that in a certain class of fractional Chern insulators, named as topological nematic states, lattice dislocations behave like topology-changing non-Abelian defects
%A second important direction is to explore conceptually novel possibilities allowed by projective non-abelian
%statistics that would not be allowed for ordinary non-abelian statistics.

%A wide class of projective non-abelian statistics can be realized in more conventional states of matter
%that have already been realized experimentally, such as superconductors, topological insulators, or Abelian
%fractional quantum Hall (FQH) states, making them a promising direction to search for non-abelian statistics.
%Therefore a second important direction is to

In this paper, we develop a more systematic understanding of the projective statistics of these extrinsic defects
and the relation between different types of extrinsic defects. We study a wide class of defects that obey projective non-abelian statistics, and that can be interpreted
as \it twist defects \rm associated with symmetries of the topological state. The examples discussed in the last paragraph can all be understood as twist defects. For example, a Chern number $2$ topological nematic state is mapped to a bilayer FQH state, with a $Z_2$ symmetry associated with exchanging the two layers. The lattice dislocation is a $Z_2$ twist defect in the sense that a quasi-particle going around the dislocation once will be acted upon by the $Z_2$ operation of exchanging the two layers. In the simplest case, a quasi-particle in one layer will end up in the other layer upon winding around the dislocation. \cite{barkeshli2010, barkeshli2012prx} Such a $Z_2$ twist defect can also be generalized to any other Abelian or non-Abelian bilayer topological states.
%in a double layer FQH state, there can be a $Z_2$ symmetry in the topological quantum numbers associated with exchanging the layers. This then allows the possibility of considering $Z_2$ twist defects such that quasiparticles encircling the defect exchange layers.\cite{barkeshli2010, barkeshli2011}
Similarly, a $Z_N$ topological state has a $Z_2$ symmetry
associated with electric-magnetic duality, allowing for the possibility of twist defects that exchange electric and magnetic quasiparticles as they encircle the defect.\cite{bombin2010, barkeshli2010, you2012,you2012b} Other examples that we will study in this paper include the twist defects associated with a certain particle-hole symmetry in FQH states,
which are equivalent to the edge defects in FQH/FQSH systems studied recently\cite{fu2009d, lindner2012,clarke2012,cheng2012, you2012,you2012b,vaezi2012}, and the $Z_3$ twist defects that can appear in triple-layer FQH or topological nematic states.
%The examples we study include the $Z_2$ twist defects in the double-layer FQH states and $Z_N$ states,
%the $Z_2$ twist defects that can appear in single-component quantum Hall states,\cite{fu2009d, lindner2012,clarke2012,cheng2012, you2012, vaezi2012} described by $U(1)_N$ Chern-Simons (CS) theory, $Z_3$ twist defects that can appear in triple-layer FQH states, and $Z_2$ twist defects that can appear in double layer non-abelian states.

We develop several complementary methods of computing the projective non-abelian braiding statistics of these twist defects. We find that in
the class of examples that we study, the state with $N$ twist defects can always be mapped -- in a certain precise sense -- to a topological state without
twist defects, but on a genus $g \propto N$ surface. Then, we find that the braiding of the twist defects realizes
adiabatic modular transformations, or Dehn twists, of the topological state on the high genus surface. This provides
a physical way to implement elements of the mapping class group of a topological state on a high genus surface.
Since adding twist defects effectively increases the genus, we refer to them as \it genons\rm.\footnote{
We would like to clarify that this name has a different meaning from the concept of \it geon \rm in the general relativity literature, which refers to a certain type of gravitational electromagnetic soliton solution.\cite{wheeler1955} However, we also note that the \it topological geon \rm studied in the quantum gravity literature\cite{sorkin1997} has some similarity to our genon, since both are related to manifold topology and the mapping class group.}

One particularly interesting example is given by genons in an Ising$\times$Ising topological state. The Ising topological state is a simple non-Abelian topological state with three topological quasiparticles, which can be realized in Kitaev's honeycomb lattice model\cite{kitaev2006}. By Ising$\times$Ising, we are referring to a bilayer state with each layer corresponding to an Ising theory. It is known that braiding particles in the Ising theory is not sufficient for universal topological quantum computation. \cite{preskillLectures} However, we will show that utilizing the braiding of twist defects in an Ising$\times$Ising theory can make the state universal, since it realizes Dehn twists of a single Ising theory on high genus surfaces\cite{bravyi200universal,freedman2006universal}. We will also demonstrate that the average degree of freedom, {\it i.e.}, the ``quantum dimension", of each twist defect in this theory is $d=2$. This provides an interesting example where a defect with
integer quantum dimension can allow for universal TQC while the host topological state is by itself non-universal.
%It was believed that no universal quantum computation is possible in a topological state of which all quasi-particles have squareroot-of-integer quantum dimensions. Therefore this example %provides a sharp distinction between extrinsic twist defects and intrinsic topological quasi-particles, while also providing a new approach to topological quantum computation.

The genons are confined in the sense that there is a long-ranged confining potential that grows with their separation. One way
to deconfine them is to gauge the symmetry they are associated with.\footnote{For $Z_2$ twist defects, we expect this can occur
physically by proliferating double-defects.} This leads to a new class of topological states,\cite{barkeshli2010, barkeshli2010twist, barkeshli2011orb}
where the defects are now intrinsic non-abelian quasiparticles. In this paper, we discuss the relation of the braid matrices between the cases
where the genons are confined, and cases where they are deconfined by gauging the associated symmetry. In the case
where the genons have qauntum dimension $\sqrt{2}$, they can be interpreted as Majorana fermions,
and gauging the symmetry does not change the braid matrix, but only makes the overall phase well-defined and universal.
In the more general situations, we find that gauging the symmetry changes the dimension of the braid matrix, so that strictly speaking the
braiding is different, though closely related. In the single-component case, we find that gauging the symmetry can even change the
quantum dimension of the non-abelian defects. These provide simple examples where projective non-abelian braiding statistics can give
braiding that is inequivalent to ordinary non-abelian statistics.

%In the case of double layer non-abelian states, of the form $G \times G$, where $G$ labels a non-abelian topological state,
%it is possible to consider $Z_2$ twist defects associated with the $Z_2$ layer exchange symmetry. We find that the quantum dimension
%of the defects is
%\begin{align}
%d_{Z_2} = D,
%\end{align}
%where $D$ is the total quantum dimension of $G$. When $G$ is the Ising topological state, the $G \times G$ state by itself is non-universal
%for TQC. However, as we discuss, the twist defects have quantum dimension $d_{Z_2} = 2$ and can be used to make the state
%universal for TQC. It has been conjectured that for a non-abelian topological state can be universal for TQC if and only if
%it has non-abelian quasiparticles with quantum dimension whose square is non-integer.\cite{rowell2009} Therefore, this provides a striking example
%of fundamentally novel behavior that can arise for projective non-abelian statistics.

The rest of this paper is organized as follows. In Sec. \ref{twistDefSec}, we introduce the notion of a twist defect in a topological state and we
discuss several examples, possible physical realizations, and we give a brief discussion of projective non-abelian statistics. In Sec. \ref{Z2twocomp}, we study in detail the braiding of $Z_2$ twist defects in ``two-component'' states, which can described by $U(1) \times U(1)$ CS theory and which include double layer FQH states and $Z_N$ topological states. We discuss the sense in which the twist defects can be thought of as ``genons'' (subsequently, we will use ``twist defect'' and ``genon'' interchangeably in this paper). In Sec. \ref{Z2onecomp}, we study $Z_2$ genons in single-component states, which are described by $U(1)_N$ CS theory, and we discuss the close relation between these genons and the ones in the two-component case. In Sec. \ref{Z2na}, we study $Z_2$ genons in decoupled, double-layer non-abelian states, and discuss the possibility of universal TQC with genons. In Sec. \ref{Z3threecomp}, we study $Z_3$ genons in three-component FQH states, which are described
by $U(1) \times U(1) \times U(1)$ CS theory; this provides the first example of projective non-abelian braiding beyond the $Z_2$ genon case.
In Section \ref{orbifoldSec}, we discuss topological states that are obtained when
the symmetry associated with the genons are gauged, and we discuss the relation between the braiding of the genons
before and after the symmetry is gauged. %In Section \ref{TQCSec}, we discuss the possibility of universal TQC with genons, and
We conclude with a discussion in Section \ref{discSec}.

\section{Twist defects in topological states}
\label{twistDefSec}

\subsection{General definition of twist defect}

A topologically ordered phase\cite{wen04, nayak2008} is generally characterized by a set of topologically non-trivial quasiparticles,
$\{\gamma_i\}$, for $i = 1, \cdots, N_{qp}$, where $N_{qp}$ is the number of quasiparticles. Below, we will briefly sketch the
topological properties of the quasiparticles. First, when two quasiparticles are observed from far away, they behave like a superposition of single quasiparticle states.
This is described by the fusion rules $\gamma_i\times \gamma_j= \sum_k N_{ij}^k\gamma_k$. Secondly, when two quasiparticles
$\gamma_i,\gamma_j$ wind around each other, a phase $e^{i\theta_{ij}^k}$ is obtained, which depends on the fusion channel
$k$. $\theta_{ij}^k$ is referred to as the braid statistics of the quasiparticles. When a particle is spinned around itself by $2\pi$, it
generically gains a nontrivial phase $e^{i\theta_i}$. $\theta_i=0$ for bosons and $\pi$ for fermions, and in general it can take
any value between $[0,2\pi)$. The braiding, fusion rules and spins need to satisfy some consistency conditions but we
will not review them here.\cite{preskillLectures} Mathematically, a topologically ordered state is characterized by a
unitary modular tensor category (UMTC). \cite{kitaev2006, preskillLectures}\footnote{Strictly speaking, a UMTC describes a topologically
ordered state of bosons. The mathematical framework for topologically ordered states of fermions is slightly different, as the spins $\theta_i$
are only defined modulo $\pi$.}

%XLnote: A illustration figure on fusion rule and braiding will be helpful here.

%
%specified for example by the  spins $\{\theta_i\}$, and a so-called modular $S$-matrix,
%which encodes the fusion rules and mutual statistics of the quasiparticles.\cite{kitaev2006,preskillLectures} Mathematically, the quasiparticles
%form the objects in a unitary modular tensor category (UMTC). The phase obtained
%by the wave function as quasiparticle $\gamma_i$ rotates by $2\pi$ is set by the spins: $e^{i\theta_i}$,
%and this is usually packaged in the T-matrix $T_{ij} = \delta_{ij} e^{i \theta_i}$.

It is possible for a topological phase to have a discrete symmetry %$G$ associated with its topological quantum numbers. An element $g\in G$
$g$ which maps a quasiparticle $\gamma_i$ to another particle $\gamma_{g(i)}$,
\begin{align}
\gamma_i \rightarrow \gamma_{g(i)}, \;\; i = 1, \cdots, N_{qp}
\end{align}
while preserving all topological properties such as fusion rules, braiding and spins.
All such symmetries form the group of automorphisms of the UMTC.
%This means that all of the topological data are invariant under the transformation
%\begin{align}
%\gamma_i \rightarrow g_{ij} \gamma_j,
%\end{align}
%for each $g \in G$. In particular,
%\begin{align}
%[S,g] = 0, \;\;\; [T,g] = 0.
%\end{align}
%$G$ is therefore the group of transformations that keep the topological quantum numbers invariant; mathematically, it is the group of automorphisms of the UMTC.
Table \ref{twistExTable} summarizes some examples. For instance,
a $Z_N$ topological state has $N^2$ quasiparticles, which can be labelled as
$(a,b)$, for $a,b = 0, \cdots, N-1$. The $(a,0)$ particles are the electric
particles, while the $(0, a)$ particles are the magnetic ones. This state has
a $Z_2 \times Z_2$ symmetry. One of the $Z_2$ symmetries is the
electric-magnetic duality $(a,b)\rightarrow (b,a)$, \it ie \rm exchanging electric and magnetic particles. The other $Z_2$ is associated
with taking $(a,b) \rightarrow (N-a, N-b)$, which takes the electric and magnetic particles to their conjugates.

If charge conservation is broken, a $1/k$-Laughlin FQH state also has a $Z_2$ symmetry, associated with
exchanging quasiparticles and quasiholes. This is beacuse the quasiparticles and quasiholes have the
same fractional statistics and yet are topologically distinct quasiparticles.

In a bilayer FQH state, there can be a $Z_2$ symmetry associated with exchanging layers. In an $N$-layer FQH state the symmetry of
permuting the layers is $S_N$, which may be broken to a smaller subgroup, such as $Z_N$.
% If we break charge conservation, for some FQH states there can be a different $Z_2$ symmetry associated with exchanging quasiparticles and quasiholes.

\begin{table*}
\begin{tabular}{ccl}
\multicolumn{3}{c}{Examples of Symmetries of Topological States} \\
\hline
Topological states&Symmetries&Transformation of quasiparticles\\
\hline
\multirow{2}{*}{$Z_N$ states} & Electric-magnetic duality $Z_2$ & $(a,b) \rightarrow (b,a)$ \\
& Particle-hole symmetry $Z_2$ &  $(a,b) \rightarrow (N-a, N-b)$ \\
$N$-layer FQH states & Layer permutation $S_N$ & $(a_1,a_2,..,a_N)\rightarrow (a_{P_1},a_{P_2},...,a_{P_N})$ \\
$1/k$-Laughlin FQH state & Particle-hole symmetry $Z_2$ & $a \rightarrow (k - a)$. \\
\hline\\
\end{tabular}
\caption{
\label{twistExTable}
Some examples of topological phases and their symmetries. $Z_N$ topological states have $Z_2 \times Z_2$
symmetry associated with both the electric-magnetic duality ( $(a,b) \rightarrow (b,a)$ ) and the particle-hole transformation taking
quasiparticles to their conjugates: $(a,b) \rightarrow (N-a, N-b)$, where $a,b = 0, \cdots, N-1$.
$N$-layer FQH states can have a symmetry associating with permuting layers. $1/k$-Laughlin states
can have a particle-hole symmetry associated with taking a quasiparticle to its conjugate, if charge conservation is broken.
}
\end{table*}

%XLnote: I have modified the table.

\begin{figure}[tb]
\centerline{
\includegraphics[width=1.5in]{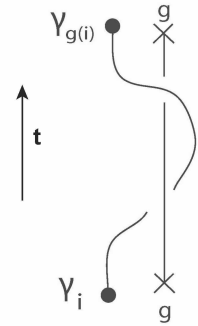}
}
\caption{
\label{qptwists}
Worldline of quasiparticle $\gamma_i$ and twist defect, labelled by $g$.
Braiding $\gamma_i$ around the twist defect changes the quasiparticle to $\gamma_{g(i)}$. The arrow indicates the time direction.
}
\end{figure}
Given a topological phase with such a symmetry, a twist defect is an extrinsic defect of the system, labelled
by the symmetry $g$, such that a quasiparticle $\gamma_i$ gets transformed to $\gamma_{g(i)}$ as
it braids around the twist defect (see Fig. \ref{qptwists}). Such twist defects generally have a non-trivial
quantum dimension and lead to topological degeneracies. Therefore, every automorphism of the UMTC
is associated with a twist defect that realizes projective non-abelian statistics.

The twist defects discussed above are related to recent discussions in the mathematical physics
literature,\cite{kitaev2012, fuchs2012, beigi2011,kapustin2010,kapustin2011} where the
mathematical theory of twist defects and boundaries between topological states is currently under investigation.

There are also twist defects that cannot be included in the definition above---those that act trivially
on the topologically distinct quasiparticles, but have a non-trivial action
on local operators. In these situations, it is possible but not guaranteed that the twist defect
will have a non-trivial quantum dimension. One example is a system with charge conservation, where the
twist defect has the effect of sending a quasiparticle to its conjugate. In an IQH state, such defects
can trap Majorana zero modes, even though the IQH state does not have any topologically non-trivial
quasiparticles. A second example is a superconductor, with the $Z_2$ symmetry that takes the
electron $c \rightarrow -c$. The vortex in $p+ip$ topological superconductors, which traps a Majorana zero mode,
can then be interpreted as a twist defect that acts on the fermions by this $Z_2$ symmetry as they encircle the vortex.
In these cases, the non-trivial quantum dimension is protected by a fermion parity symmetry. 

\subsection{Physical Realizations}

Recently, twist defects have been shown to occur in a number of different physical realizations. These include
dislocations in topological nematic states realized in FCI with higher Chern number $C > 1$\cite{barkeshli2012prx},
at certain junctions in gated bilayer FQH states\cite{barkeshli2012prop}, dislocations in exactly solvable $Z_N$
topologically ordered models\cite{bombin2010, you2012, kitaev2012,you2012b}, and  superconductor-ferromagnet domain walls at the
edge of 2+1d TIs and FTIs.\cite{fu2009d,lindner2012,clarke2012,cheng2012}

The topological nematic state realization of the twist defects\cite{barkeshli2012prx} is based on the Wannier state representation of FCI\cite{qi2011}. This representation established a mapping between one-dimensional Wannier states of an FCI system and the Landau level wavefunctions in the Landau gauge in an ordinary FQH system. By such a mapping, a Chern number $C=1$ band is mapped to a Landau level. The validity of the Wannier state representation approach has been confirmed in $C=1$ states with filling $1/2$ and $1/3$\cite{wu2012,scaffidi2012}. FCI with $C>1$ bands have also been studied in analytic and numerical works\cite{barkeshli2012prx,wang2011c,wang2012,sterdyniak2012}. In the Wannier state representation, a band with Chern number $C > 1$ is mapped to a $C$-layer quantum Hall system\cite{barkeshli2012prx}. In contrast to ordinary multi-layer quantum Hall systems, the different effective layers in this system are related by lattice translations, which implies that lattice dislocations can act as twist defects: as a reference point encircles the
dislocation, it is translated by the Burgers vector of the dislocation and may effectively belong to a different layer of the effective FQH system.
%XLnote: This section needs to be expanded to emphasize the topological nematic state physics.
%MBnote: I haven't done this -- not sure how much to explain.

The realization of twist defects as lattice dislocations in the $Z_N$ models is similar.\cite{bombin2010, you2012} In some
lattice models of the $Z_N$ states, the electric particles ($Z_N$ charges) belong to the even sublattice, while the
magnetic particles ($Z_N$ vortices) belong to the odd sublattice. A lattice dislocation can create a situation
where an excitation that encircles it starts in the even sublattice and ends in the odd sublattice, which therefore implies
that the dislocation can act as a twist defect whereby electric and magnetic particles are exchanged upon encircling it.

As we will explain in subsequent sections, the superconductor-ferromagnet domain walls at junctions of IQH/FQH states
can be viewed as twist defects in several different ways: they can be viewed as twist defects associated with a $Z_2$
particle-hole symmetry of the Abelian QH states, or they can be mapped onto double layer FQH states, in which case
they are associated with the $Z_2$ layer exchange symmetry.

\subsection{Projective non-abelian braiding statistics}

The twist defects are \it extrinsic \rm defects of the topological phase, not finite-energy, intrinsic excitations.
Therefore, separating twist defects will typically cost an energy that grows with the distance between them, either
logarithmically or linearly, depending on whether the associated symmetry is a continuous symmetry of the state.
This is similar to the vortices in topological superfluids (and thin-film topological superconductors), which
also have a logarithmic energy cost associated
with separating vortex/anti-vortex pairs. In these cases, we say that the defects are \it confined \rm.

Since the defects can have a non-trivial quantum dimension, braiding them can lead to non-abelian braiding statistics.
However, since well-separated twist defects still have an energy cost that grows with the separation, the overall phase
of the braiding statistics is not well-defined.
To see this, observe that simply moving one quasiparticle in a small circle without encircling any other quasiparticle still accumulates a phase
from the non-negligible change in energy of the state during the process. Since the overall phase of the statistics is not well-defined,
these twist defects form a \it projective \rm representation of the braid group. In contrast, in a conventional non-abelian
state, the non-abelian quasiparticles are finite energy, deconfined excitations, and the overall phase of the braiding is topological.
Therefore, the braiding of non-abelian quasiparticles in a true non-abelian state forms a linear representation of the braid group.

The quasiparticles of a true non-abelian state are subject to the mathematical constraints of a UMTC. However, the twist defects, since they
form a projective representation of the braid group and are not intrinsic quasiparticles, are described by a different mathematical theory
(see \it eg \rm Ref. \onlinecite{freedman2011}), and
in principle are not constrained in the same way. This opens the possibility of fundamentally novel behavior. The difference between twist defects with projective non-Abelian statistics and intrinsic topological quasi-particles will be discussed further in Sec. \ref{orbifoldSec}, and the application to UMTC will be discussed in Sec. \ref{Z2na}.

\section{$Z_2$ twist defects in two-component Abelian states}
\label{Z2twocomp}

In this section, we concentrate on the properties of $Z_2$ twist defects in Abelian states that
can be described by $U(1) \times U(1)$ CS theory. This includes two-component FQH
states\cite{wen04} and $Z_N$ topological states. The Lagrangian is given in terms of two $U(1)$ gauge fields,
$a$ and $\tilde{a}$:
\begin{align}
\label{bilayerCS}
\mathcal{L} = \frac{m}{4\pi} (a \partial a + \tilde{a} \partial \tilde{a}) + \frac{l}{4\pi} (a \partial \tilde{a} + \tilde{a} \partial a),
\end{align}
where $m$ and $l$ are any integers and $a \partial a \equiv \epsilon^{\mu \nu \lambda} a_\mu \partial_\nu a_\lambda$.
To describe $Z_N$ topological states, we set $m = 0$ and $l = N$ above. This theory encodes the
topological properties of these topological states. As we review below, the $Z_2$ twist defects in
these states are non-Abelian defects, carrying a quantum dimension $\sqrt{|m - l|}$.\cite{barkeshli2010, barkeshli2012prx}
Based on the field theory description in Ref. \onlinecite{barkeshli2010, barkeshli2012prx}, we will derive the braid matrix of the $Z_2$ twist defects.

\subsection{Bulk geometrical picture}
\label{bulkZ2}

The twist defects introduced above are point defects: far away, no local operator can distinguish their
presence. However, for the purpose of understanding the behavior of the twist defects,
it is helpful to imagine that the twist occurs along a single branch cut that connects them. This is
similar to supposing that the phase winding of a vortex in a superconductor is all localized to a single
cut connecting a vortex/anti-vortex pair.
\begin{figure}
\centerline{
\includegraphics[width=2.5in]{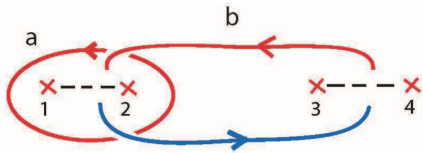}
}
\caption{A pair of twist defects induces two distinct non-contractible loops, labelled $a$ and $b$.
For twist defects in $(mm0)$ states, these loops effectively cross only once, leading to a magnetic algebra for the
quasiparticle loop operators. Here, the branch cut indicated by the dash line connecting two dislocations is
merely a gauge choice, which is similar to supposing that the phase winding of a vortex in a superconductor is
all localized to a single cut connecting a vortex/anti-vortex pair.
\label{2loops}
}
\end{figure}

The twist defects introduce new non-contractible loops into the system, along which the quasiparticles can propagate.
For example, with two pairs of $Z_2$ twist defects on a sphere, there are two distinct non-contractible
loops (Fig. \ref{2loops}). The key feature of the twist defects is that their presence yields a non-trivial algebra
for quasiparticle loop operators corresponding to these non-contractible loops. For example, consider the
case of two decoupled $1/m$ Laughlin FQH layers (denoted $(mm0)$ states), where the twist defect exchanges the layers.
The two loops $a$ and $b$ effectively cross only once, because a quasiparticle goes from one layer to another as it passes
through the branch cut, and the quasiparticles in two different layers have trivial mutual braid statistics. This leads to the magnetic algebra
\begin{align}
W(a)W(b) = W(b) W(a) e^{2\pi i/m},
\end{align}
where $W(C)$ is the operator that tunnels a Laughlin quasiparticle around the loop $C$. The explicit expression of $W(C)$ is given later in Eq. (\ref{eq:Wilsonloopdef}). The ground states form
an irreducible representation of this quasiparticle loop algebra, which in this example is $|m|$-dimensional. Therefore
we can conclude that the twist defects have a non-trivial quantum dimension.

\begin{figure}
\centerline{
\includegraphics[width=3in]{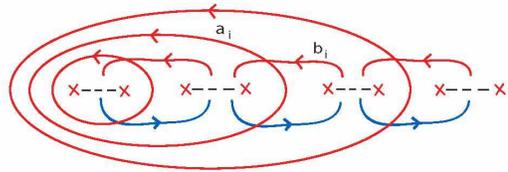}
}
\caption{$n$ pairs of twist defects induces $2(n-1)$ distinct non-contractible loops,
$a_i, b_i$, for $i = 1,\cdots, n-1$. For twist defects in the $(mm0)$ states, the quasiparticle
loop operators give rise to $n-1$ copies of the magnetic algebra, leading to $|m|^{n-1}$ topologically
degenerate states.
\label{manyLoops}
}
\end{figure}

Proceeding with the example of the $(mm0)$ states, with $n$ pairs of twist defects on a sphere,
we can define $2(n-1)$ non-contractible loops (see Fig. \ref{manyLoops}),
$a_i$, $b_i$ for $i = 1, \cdots, n-1$, such that $a_i$ and $b_j$ effectively cross exactly once
if $i = j$, and do not cross otherwise.

Using these non-contractible loops, the quasiparticle loop algebra now becomes
\begin{align}
W(a_i) W(b_j) = W(b_j) W(a_i) e^{\delta_{ij} 2\pi i /m}.
\end{align}
Thus we have $n-1$ copies of the magnetic algebra, and therefore
a finite dimensional irreducible representation of dimension $|m|^{n-1}$.
This shows that the quantum dimension of each twist defect in the $(mm0)$ states is $\sqrt{|m|}$.
\begin{figure}
\centerline{
\includegraphics[width=3in]{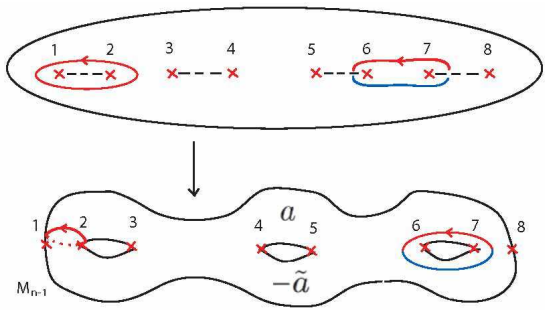}
}
\caption{
\label{disloc}
The $U(1) \times U(1)$ CS theory with $n$ pairs of dislocations on a sphere can be mapped
to a $U(1)$ CS theory on a genus $g = n-1$ surface, $M_{n-1}$. $M_{n-1}$ consists of two copies
of the original space. A new $U(1)$ gauge field, $c$, is defined on $M_{n-1}$, such that
$c = a$ on the top half, and $c = -\tilde{a}$ on the bottom half of $M_{n-1}$.\cite{barkeshli2010}
}
\end{figure}

It is possible to derive the above results more concretely by starting with the field theory (\ref{bilayerCS}).
Here we briefly review the field theory in the presence of twist defects developed in Ref. \onlinecite{barkeshli2010}.
In the $U(1) \times U(1)$ CS theory, a pair of $Z_2$ twist defects can be modelled as a pair of points,
connected by a branch cut $\gamma$ such that at the branch cut, the two gauge fields $a$ and $\tilde{a}$ obey
twisted boundary conditions. Defining $A = \left( \begin{matrix} a & 0 \\ 0 & \tilde{a} \end{matrix} \right)$,
this means
\begin{align}
\lim_{p \rightarrow p_0^{\pm}} A(p) = \lim_{p \rightarrow p_0^{\mp}} \sigma_x A(p) \sigma_x,
\end{align}
for every point $p_0$ on $\gamma$.
\footnote{To be more precise, the condition above is not in a gauge invariant form and we should require it to be satisfied in a certain gauge choice.}
The limit $p \rightarrow p_0^{+(-)}$ means that the limit is taken
approaching one particular side (or the other) of $\gamma$. This ensures that quasiparticles encircling
a twist defect get transformed by the $Z_2$ action, $\sigma_x = \left( \begin{matrix} 0 & 1 \\ 1 & 0 \end{matrix} \right)$.

In the presence of $n > 1$ pairs of twist defects, it is possible to consider a single
$U(1)$ gauge field $c$ on a doubled space, $M_{n-1}$, where $M_{n-1}$ is a genus
$g = n-1$ surface (see Fig. \ref{disloc}). As shown in Fig. \ref{disloc}, $c = a$
for points in the top half of $M_{n-1}$ and $c = -\tilde{a}$ for points in the bottom half
of $M_{n-1}$. The original $U(1) \times U(1)$ CS theory (\ref{bilayerCS})
can then be shown to equivalent to a $U(1)_{m-l}$ CS theory on $M_{n-1}$:\cite{barkeshli2010}
\begin{align}
L &= \frac{m-l}{4\pi} \int_{M_{n-1}} c \partial c.
\end{align}
The quasiparticle loop operators $W(C)$ defined earlier in the context of the $(mm0)$ states are written
in the field theory as
\begin{align}
W(C) = \mathcal{P} e^{i \oint_C c \cdot dl}.\label{eq:Wilsonloopdef}
\end{align}
Such a theory has a ground state degeneracy of $|m-l|^{n-1}$, which shows that the $Z_2$ twist defects
have a quantum dimension $\sqrt{|m-l|}$.\cite{barkeshli2010, barkeshli2012prx}
\begin{figure*}
\centerline{
\includegraphics[width=6in]{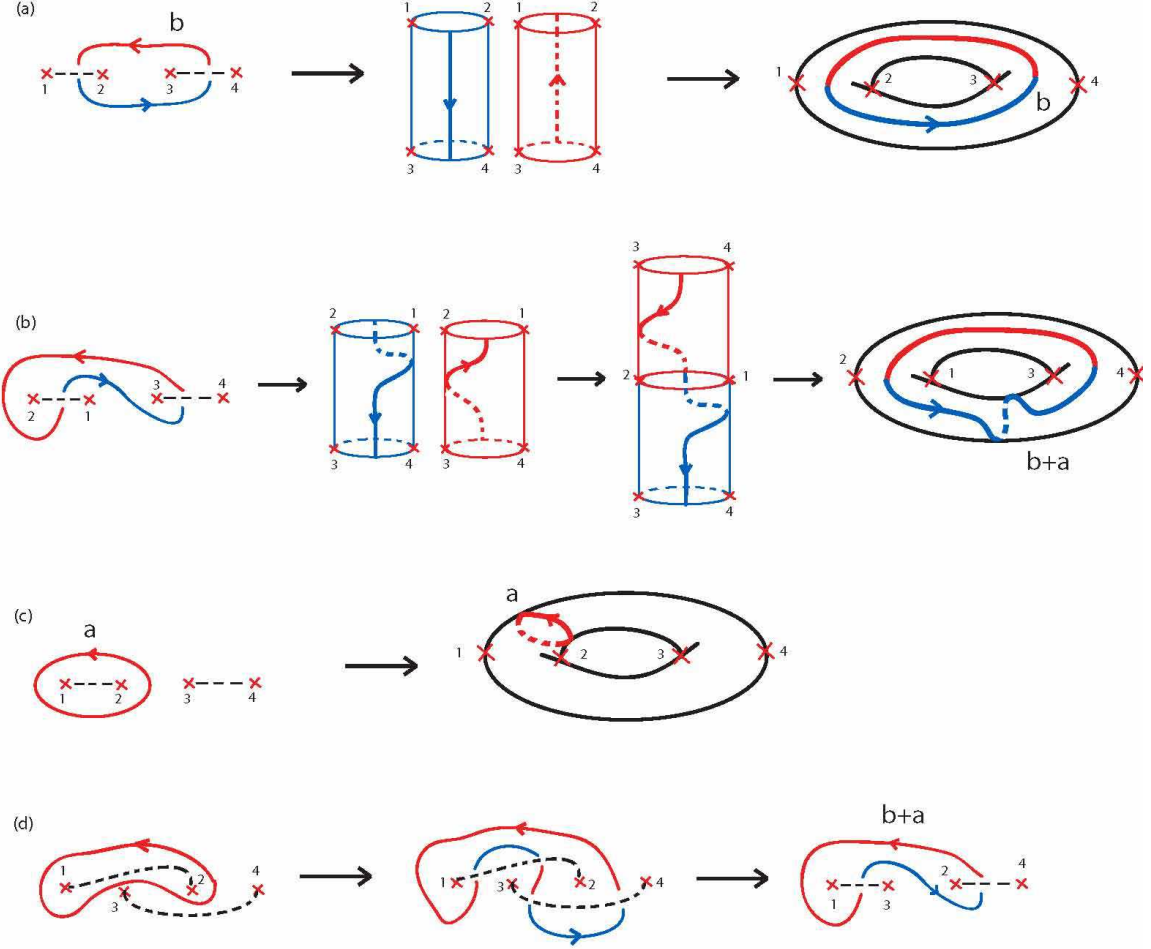}
}
\caption{(a) A loop that encloses the twist defects labelled $2$ and $3$ is mapped to the $b$ cycle of the torus.
(b) Effect of a counterclockwise exchange of $1$ and $2$. By following the effect on the non-contractible loop, we
see that in terms of the genus $g$ surface, it has the effect of a Dehn twist along an $a$ cycle. Thus, the original
$b$ loop becomes the loop $b+a$ after counterclockwise exchange of $1$ and $2$.
(c) A loop that encloses a pair of twist defects connected by a branch cut is an $a$-cycle of the genus $g$ surface.
(d) Effect of a clockwise exchange of $2$ and $3$. We see that the $a$ loop gets mapped to the $a + b$ loop. Thus
the clockwise exchange of $2$ and $3$ is equivalent to a Dehn twist along the $b$ loop.
\label{loops}
}
\end{figure*}
If the $n$ pairs of twist defects were placed on a genus $g$ surface instead of a sphere, then the ground
state degeneracy would be given by $|m^2 - l^2|^g \times |m-l|^{n-1}$, where the factor $|m^2 - l^2|^g $
comes from the ``bare'' degeneracy of the manifold on which the twist defects are placed.

Equipped with the field theory description, we can now derive the non-abelian statistics of the $Z_2$ twist defects.
The explicit mapping to the high genus surface is particularly useful for understanding of the braiding properties.
First, we exchange the twist defects on the sphere, and then
we can visualize how the non-contractible loops transform under the exchange. From Fig. \ref{loops}a and
\ref{loops}b, we can see that under a counterclockwise exchange $B_{12}$ of the defects labelled $1$ and $2$,
the loops transform as
\begin{align}
B_{12}: b \rightarrow b + a.
\end{align}
From Fig. \ref{loops}c and \ref{loops}d, we see that under clockwise exchange of the defects labelled
$2$ and $3$,
\begin{align}
B_{23}^\dagger: a \rightarrow a +b .
\end{align}
Therefore, we see explicitly that the braiding of the twist defects correspond to adiabatic modular transformations (Dehn twists)
in the effective genus $g$ surface. The adiabatic Berry phase associated with braiding twist defects is then
given by non-abelian adiabatic Berry phases associated with Dehn twists, which were computed
in \Ref{wen1990naberry} for $U(1)$ CS theory. Note that as discussed in
\Ref{wen1990naberry}, these non-abelian adiabatic Berry phases are well-defined only up to the overall phase of the matrix,
reflecting the fact that the non-abelian braiding statistics are \it projective\rm.

%\begin{figure}
%\centerline{
%\includegraphics[width=3in]{dehntwists.eps}
%}
%\caption{Labelling of some of the basic non-contractible loops on a genus $g$ surface.
%$a_i$, for $i =1,\cdots, g+1$, $b_i$ for $i = 1,\cdots, g$ and $c_i$ for $i = 1,\cdots, g-1$.
%\label{dehntwists}
%}
%\end{figure}

Using the relation to Dehn twists, we can now immediately obtain the braid matrices.
Consider two pairs of twist defects on a sphere, so that we are mapped to a $U(1)_{m-l}$
CS theory on a torus. Let us consider a basis that consists of wrapping the quasiparticles
around the $b$ direction of the torus:
\begin{align}
|n \rangle, \;\;\; n = 0, \cdots, |m-l| - 1.
\end{align}
This corresponds to diagonalizing $W(a)$, while $W(b)$ acts as a raising operator:
\begin{align}
W(b)|n \rangle &= |n + 1 \text{ mod } |m-l| \rangle,
\nonumber \\
W(a) |n \rangle &= e^{2\pi i n /(m-l)} |n \rangle.
\end{align}
Dehn twists along a non-contractible loop $C$ will be denoted by $U_{C}$.
Based on the previous discussion,
\begin{align}
B_{12} = U_a, \;\;\; B_{23}^\dagger = U_b.
\end{align}
It was found\cite{wen1990naberry} that $U_a$ are diagonal in the above basis, and given by
\begin{align}
\label{Z2Dehn1}
U_a |n \rangle = e^{i\theta} e^{i \pi n^2/(m-l) + i \pi n (m-l)} |n \rangle,
\end{align}
where the overall phase $e^{i\theta}$ depends on details of the path, not only on its topology.
In order to compute $U_b$, we observe that
\begin{align}
\label{Z2Dehn2}
U_b^\dagger = S^\dagger U_a S,
\end{align}
where $S$ is the modular $S$ matrix, which exchanges the $a$ and $b$ cycles:
\begin{align}
S: \;\;\; a \rightarrow b, \;\;\; b \rightarrow -a.
\end{align}
The adiabatic modular transformation associated with $S$ is given by\cite{wen1990naberry}
\begin{align}
S_{\alpha \beta} = \frac{1}{\sqrt{|m-l|}}e^{-2\pi i \alpha \beta/(m-l)}.
\end{align}
Therefore:
\begin{align}
\label{Z2Dehn2b}
(U_b^\dagger)_{\alpha \delta} = \frac{e^{i\phi}}{|m-l|}\sum_{\beta=0}^{|m-l|-1} e^{i \pi \beta (m-l) +i \pi (\beta^2 + 2\beta(\alpha - \delta))/(m-l)},
\end{align}
where the overall phase $e^{i\phi}$ again depends on non-universal details of the path.

Therefore, we find that in the presence of the twist defects, there is a group of projective non-abelian braiding statistics
associated with the braiding of the defects, along with the braiding of the quasiparticles around the defects. These are
generated by the unitary operators
\begin{align}
\{B_{i, i+1}, W(a_i), W(b_i)\}.
\end{align}

The generalization of the braid matrices to $n > 2$ pairs of twist defects
is straightforward, since for the $U(1)_{m-l}$ CS theory, the states associated with
each handle of the genus $g = n-1$ surface are independent.

We see that the effect of the $n$ twist defects is to introduce a set of non-trivial quasiparticle loop operators that satisfy
the same algebra as the quasiparticle loop operators of a $U(1)_{m-l}$ CS theory on a genus $n-1$ surface, and
that braiding of the twist defects is directly related to elements of the mapping class group of this theory on the high
genus surface. In order to emphasize the connection between the twist defects and the high genus surface, we
refer to the twist defects as genons. All of the twist defects considered in this paper have this property, and therefore
in this paper we will use ``twist defects'' and ``genons'' interchangeably.

\subsection{An example: Majorana braiding as Dehn twist of $1/2$ Laughlin state}

The simplest case where genons are non-abelian is for the case $| m - l | = 2$, in which case
they have quantum dimension $\sqrt{2}$. We therefore expect that a Majorana zero mode is localized
at the defect, and the braiding statistics should be associated with the braiding of Ising anyons.

Let us consider the case where we have $n = 2$ pairs of genons on a sphere, which maps us to a genus
$g = 1$ surface. In this case, there are $|m-l|^g = 2$ ground states.
The braiding of the genons, given by the Dehn twists are:
\begin{align}
U_a = e^{i\theta} \left(\begin{matrix}
1 & 0 \\
0 & i
\end{matrix} \right) ,\;\;
U_b = e^{i\phi}\frac{1}{2}\left(\begin{matrix}
1+i & 1-i\\
1-i & 1+i
\end{matrix} \right).
\end{align}
Observe that $U_a$ is, up to an overall phase, the same as the braid matrix of Ising anyons!\cite{nayak2008}
In other words, Dehn twists in the $U(1)_2$ CS theory are equivalent to projective Ising braiding statistics, due to
this profound relation to $Z_2$ genons.
%Note that both in this case, and in the case of topological superconductors, the adiabatic braiding
%of the Majorana fermions is well-defined only up to an overall phase.\footnote{In the topological
%superconductor examples, this is because there is a long-range logarithmic interaction between
%the vortices that host the Majorana zero modes.}

In addition to the braiding of the genons, we can also braid quasiparticles around genons to get:
\begin{align}
W(a) = \left(\begin{matrix}
1 & 0 \\
0 & -1 \\
\end{matrix} \right), \;\;\;
W(b) = \left(\begin{matrix}
0 & 1 \\
1 & 0 \\
\end{matrix} \right).
\end{align}

\subsection{1+1D edge CFT  picture}

It is helpful to derive the above results in a different way using the one-dimensional chiral Luttinger liquid edge
theory. This will give us a different perspective on how to compute the topological degeneracy and braiding statistics,
it will connect to the more standard zero mode analysis used in other contexts,\cite{alicea2012review} and it will help give a different protocol
for carrying out the braiding. The latter will be useful depending on the physical realization of the genons.
\begin{figure}
\centerline{
\includegraphics[width=3.5in]{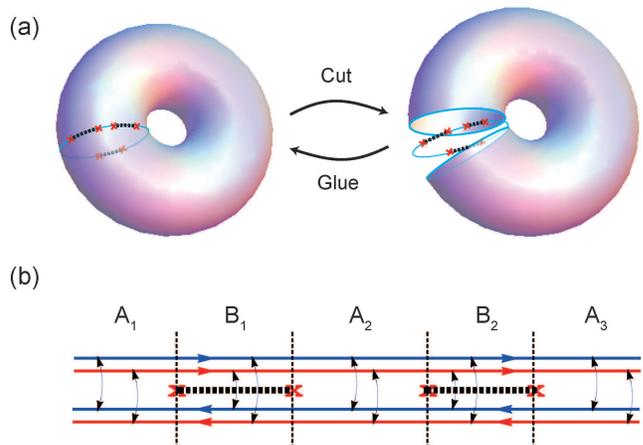}
}
\caption{The edge state understanding of the topological degeneracy.
(a) The genons are oriented along a single line, and then the system is cut along the line, yielding
gapless counterpropagating edge states along the line. The original topological state is obtained from gluing the the
system back together by turning on appropriate inter-edge tunneling terms.
(b) Depiction of the two branches (red and blue) of counter-propagating edge excitations. The arrows between the edge states
indicate the kinds of electron tunneling terms that are added. Away from the genons, in the
$A$ regions, the usual electron tunneling terms involving tunneling between the same layers, $\Psi_{eRI}^\dagger \Psi_{eLI} + H.c.$,
are added. In the regions including the branch cuts separating the genons, twisted tunneling terms are
added: $\Psi_{eR1}^\dagger \Psi_{eL2} + \Psi_{eR2}^\dagger \Psi_{eL1} + H.c.$.
\label{fig1sup} }
\end{figure}
We will start with a brief review of the field theoretic analysis of the $1+1$ edge CFT in Ref. \onlinecite{barkeshli2012prx}.
First, we align all of the defects along a single line, and we cut the system along the line. Then, we have two counter-propagating
chiral Luttinger liquid theories, with the left-moving one localized on one edge, and the right-moving one on the other edge.
The edge theory for $U(1) \times U(1)$ CS theory described by a generic $K$-matrix is given by the action \cite{wen1990edge,wen1992}
\begin{align}
\label{edgeAc}
S_{edge} = \frac{1}{4\pi} \int dx dt[ K_{IJ} \partial_t \phi_{LI} \partial_x \phi_{LJ} - V_{IJ} \partial_x \phi_{LI} \partial_x \phi_{LJ} ],
\end{align}
where in our case $K = \left( \begin{matrix} m & l \\ l & m \end{matrix} \right)$.
$\phi_{LI}$ denotes left-moving chiral bosons for $I = 1, \cdots, \text{dim } K$.\footnote{Strictly speaking, $\phi_{LI}$ is
left-moving only if the eigenvalues of $K$ are all positive. Otherwise it is either right-moving or a combination of
right- and left-moving modes. Nevertheless, we will continue to use the subscripts $L$ or $R$ even in these more general cases.}
Here and below the repeated indices $I,J$ are summed. The field $\phi_{LI}$ is a compact
boson field with radius $R = 1$:
\begin{align}
\phi_{LI} \sim \phi_{LI} + 2\pi.
\end{align}
Quantizing the theory in momentum space yields \cite{wen04}
\begin{align}
[\partial_x \phi_{LI} (x), \phi_{LJ}(y)] = -i2\pi K^{-1}_{IJ} \delta (x - y).
\end{align}
Integrating the above equation gives:
\begin{align}
[\phi_{LI}(x), \phi_{LJ}(y)] = -i\pi K^{-1}_{IJ} sgn(x-y) \label{1dCommutatorL}
\end{align}
The electric charge density associated with $\phi_{LI}$ is given by
\begin{align}
\rho_{LI} = \frac{1}{2\pi} \partial_x \phi_{LI},
\end{align}
and the $I$th electron operator is described by the vertex operator
\begin{align}
\Psi_{eLI} = e^{i K_{IJ} \phi_{LJ}} .
\end{align}
Note that normal ordering will be left implicit (\it ie \rm $e^{i K_{IJ} \phi_{LJ}} \equiv : e^{i K_{IJ} \phi_{LJ}} :$).
If we consider the FQH state on a cylinder, we will have a left-moving chiral theory on one edge, and a right-moving
chiral theory on the other edge. For the right-moving theory, the edge action is
\begin{align}
\label{RedgeAc}
S_{edge} = \frac{1}{4\pi} \int dx dt [-K_{IJ} \partial_t \phi_{RI} \partial_x \phi_{RJ} - V_{IJ} \partial_x \phi_{RI} \partial_x \phi_{RJ} ] ,
\end{align}
so that
\begin{align}
[\phi_{RI}(x), \phi_{RJ}(y)] = i\pi K^{-1}_{IJ} sgn(x-y), \label{1dCommutatorR}
\end{align}
the charge is
\begin{align}
\rho_{RI} = \frac{1}{2\pi} \partial_x \phi_{RI},
\end{align}
and the electron operator is
\begin{align}
\Psi_{eRI} = e^{-i K_{IJ} \phi_{RJ}} .
\end{align}
When $m$ is odd, the electron operators are fermionic, so we need to ensure that
$\Psi_{eRI}(x) \Psi_{eLJ}(y) = e^{i \pi K_{IJ}} \Psi_{eLJ}(y) \Psi_{eRI}(x)$. This can be done by introducing
commutation relations:
\begin{align}
[\phi_{LI}(x), \phi_{RJ}(y)] = i \pi K^{-1}_{IJ}. \label{1dZ2Klein}
\end{align}

Now suppose we have $n$ pairs of genons, such that going around each genon exchanges the
two layers. Each pair of genons is separated by a branch cut.
Let us align all the genons, and denote the regions without a branch cut as $A_i$, and the
regions with a branch cut as $B_i$ (Fig. \ref{fig1sup}A). Now imagine cutting the system along this line, introducing counterpropagating
chiral edge states. The gapped system with the genons can be understood by introducing different
electron tunneling terms in the $A$ and $B$ regions (Fig. \ref{fig1sup}B):
\begin{align}
\label{tunnL}
\delta \mathcal{H}_{t} =
\frac{g}{2} \left\{ \begin{array}{ccc}
\Psi_{eL1}^\dagger \Psi_{eR1} + \Psi_{eL2}^\dagger \Psi_{eR2}  + H.c & \text{ if } & x \in A_i \\
\Psi_{eL1}^\dagger \Psi_{eR2} + \Psi_{eL2}^\dagger \Psi_{eR1} + H.c  & \text{ if } & x \in B_i \\
\end{array} \right.
\end{align}
Introducing the variables
\begin{align}
\phi_1 &= \phi_{L1} + \phi_{R1}\nonumber\\
\phi_2 &= \phi_{L2} + \phi_{R2}\nonumber\\
\tilde{\phi}_1 &= \phi_{L1} + \phi_{R2}\nonumber\\
\tilde{\phi}_2&= \phi_{L2} + \phi_{R1}\label{phitilde}
\end{align}
we rewrite (\ref{tunnL}) as
\begin{align}
\delta \mathcal{H}_{t} =
g \left\{ \begin{array}{ccc}
\sum_I \cos(K_{IJ} \phi_J) & \text{ if } & x \in A_i \\
\sum_I \cos( K_{IJ} \tilde{\phi}_J) & \text{ if } & x \in B_i \\
\end{array} \right.
\end{align}
It is helpful to rewrite the above as
\begin{align}
\label{2compTunn}
\delta \mathcal{H}_{t} =
g \cos\left( \frac{m+l}{2} \phi_+\right)  \left\{ \begin{array}{ccc}
\cos( \frac{m-l}{2} \phi_-) & \text{ if } & x \in A_i \\
\cos( \frac{m-l}{2} \tilde{\phi}_-) & \text{ if }& x \in B_i\\
\end{array} \right.
\end{align}
where $\phi_{\pm} = \phi_1 \pm \phi_2$, and $\tilde{\phi}_\pm = \tilde{\phi}_1 \pm \tilde{\phi}_2$.
One way to understand the topological degeneracy was explained in Ref. \onlinecite{barkeshli2012prx}.
In the absence of the twist defects, there are $|(m+l)(m-l)|$ states, associated with the
distinct eigenvalues of $e^{i(\phi_1 \pm \phi_2)}$, which are given by $e^{2\pi i p_{\pm}/(m\pm l)}$, for
$p_{\pm}$ integers. Physically, $e^{i\phi_I}$ corresponds to a quasiparticle
tunneling process, where a quasiparticle from the $I$th layer is annihilated at one edge, tunnels
around the torus, and is created at the other edge. In the presence of the $n$ pairs of twist
defects, the eigenvalue of $e^{i\phi_+}$ is globally pinned everywhere (see eq. (\ref{2compTunn}) ) while $e^{i\phi_-}$ can take
$|m-l|$ different values in each of the $A_i$ regions. The operator $e^{i(\tilde{\phi}_1 - \tilde{\phi}_2)}$ is unphysical and not
gauge-invariant,\cite{barkeshli2012prx} so we cannot label the states by its eigenvalues in the $B$ regions.
This yields a total of $|(m+l)(m-l)^n|$ states and agrees with the bulk calculation of Sec. \ref{bulkZ2},
with the extra factor of $|m^2 - l^2|$ due to the fact that in this case the defects were placed on a torus
instead of a sphere.

\begin{figure}
\centerline{
\includegraphics[width=3in]{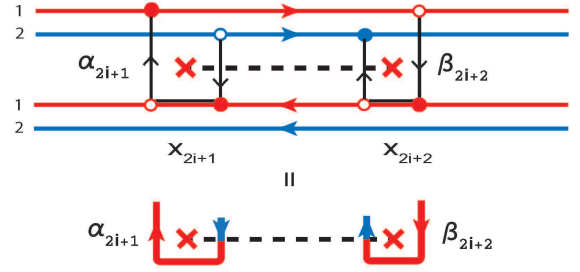}
}
\caption{Physical meaning of the zero mode operators $\alpha_i$ and $\beta_i$. In subsequent
figures, we will use the bottom figure, which is more precisely described by the top figure.
\label{zeroModes}
}
\end{figure}

Now we can use the $1+1$D edge theory to understand the braiding statistics of the defects. 
In order to calculate the braiding and to understand the connection to zero modes localized at the defects,
we consider quasiparticle tunneling operators near each genon:
\begin{align}
\alpha_{2i-1} &= e^{i\phi_1(x_{A_i}) } e^{ -i \tilde{\phi}_1(x_{B_i})},
\nonumber \\
\beta_{2i-1} &= e^{i \phi_2(x_{A_i})} e^{- i \tilde{\phi}_2(x_{B_i})},
\nonumber \\
\alpha_{2i} &= e^{i \tilde{\phi}_2 (x_{B_i} )} e^{- i\phi_2 (x_{A_{i+1}})},
\nonumber \\
\beta_{2i} &= e^{i \tilde{\phi}_1 (x_{B_i} )}  e^{-i \phi_1(x_{A_{i+1}})},
\end{align}
for $i = 1, \cdots, n$, where $x_{A_i}$ and $x_{B_i}$ are the midpoints of the $A_i$ and $B_i$ regions, respectively.
Physically, these operators are quasiparticle tunneling operators, projected
onto the ground state subspace where the fields are constant within each region. Fig. \ref{zeroModes}
displays the quasiparticle tunneling process described by these operators.
The operators $\alpha_i$ and $\beta_i$ are zero modes:
\begin{align}
[H, \alpha_i] = [H, \beta_i ] = 0,
\end{align}
where $H$ is the edge Hamiltonian including the tunneling terms (\ref{tunnL}).
In the ground state subspace, as discussed above, the value of $e^{i (\phi_1 + \phi_2)}$ is pinned everywhere. Thus:
\begin{align}
\label{c2indZM}
\alpha_{2i+1} \beta_{2i+1} &= 1 , \;\;\; &\alpha_{2i} \beta_{2i} = e^{\frac{2\pi i}{m-l}}.
\end{align}
Therefore, there is a single independent zero-mode operator, $\alpha_i$, which satisfies an algebra:
\begin{align}
\alpha_n \alpha_{n+k} &= \alpha_{n+k} \alpha_n e^{i 2\pi/(m-l)},
%\nonumber \\
%\beta_n \beta_{n+k} &= \beta_{n+k} \beta_n e^{-2\pi i/(m-l)},
%\nonumber \\
%\alpha_n \beta_{n+k} &= \beta_{n+k} \alpha_n e^{2\pi i /(m-l)},
%\nonumber \\
%\beta_n \alpha_{n+k} &= \alpha_{n+k} \beta_n e^{2\pi i/(m-l)},
\end{align}
where $k  > 0$. For $ |m - l| = 2$, $\alpha_i$ are Majorana fermions. For $|m-l| > 2$, this can be viewed
as a generalization of the Majorana fermion algebra, which is usually referred to as a $Z_{|m-l|}$
parafermion algebra.\cite{zamolodchikov1985}

Alone, the zero mode operators $\alpha_i$ are not physical gauge invariant operators, as they correspond to the
quasi-particle motion along an open path. However, proper combinations of them are physical gauge invariant operators. For example,
\begin{align}
W_1 ^\dagger (a_{i+1}) = \alpha_{2i+1} \beta_{2i+2}, \;\;\; W_2 ^\dagger (a_{i+1}) = \beta_{2i+1} \alpha_{2i+2}
\end{align}
describe quasiparticles from layer 1 and 2, respectively, tunneling around the pair of genons $2i+1$, $2i+2$.
Similarly,
\begin{align}
W_1^\dagger (b_{i}) = \alpha_{2i} \beta_{2i+1}, \;\;\; W_2^\dagger (b_{i+1}) = \beta_{2i} \alpha_{2i+1},
\end{align}
where $a_i$ and $b_i$ are as shown in Fig. \ref{manyLoops}.
Using the algebra of these quasiparticle tunneling operators in the edge theory, we can obtain the algebra
of the quasiparticle loop operators
\begin{align}
W_1(a_i) W_1(b_i) = W_1(b_i) W_1(a_i) e^{2\pi i /(m-l)}.
\end{align}
Note that from (\ref{c2indZM}), it follows that $W_1(C) \propto W_2^\dagger(C)$.
It is useful to note that
\begin{align}
\alpha_{2i+1}^{m-l} &= 1, \;\; \alpha_{2i}^{m-l} = (-1)^{m-l-1},
\end{align}
so that
\begin{align}
W_I^{m-l}(a_i) = W_I^{m-l}(b_i) = 1.
\end{align}
The irreducible representation of this algebra contains $|m-l|^{n-1}$ states. This is simply the
1+1D edge CFT understanding of the bulk geometric construction described in Section \ref{bulkZ2}.
\begin{figure}
\centerline{
\includegraphics[width=3.6in]{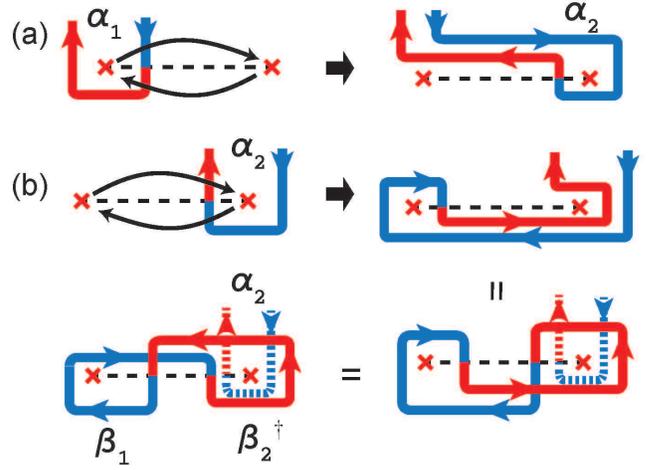}
}
\caption{Effect of a clockwise braid on the zero modes $\alpha_1$ and $\alpha_2$.
(a) $\alpha_1 \rightarrow \alpha_2$. (b) $\alpha_2$ gets transformed to a combination of two
loop, $\beta_1 \beta_2^\dagger$, and $\alpha_2$. Therefore, $\alpha_2 \rightarrow \alpha_2 \alpha_1^\dagger \alpha_2$,
where recall $\beta_i  = \alpha^\dagger_i$. As explained in the text, there is an ordering ambiguity between
$\alpha_2$ and $\alpha_1^\dagger \alpha_2$ that is fixed with an additional constraint.
\label{1dZ2exchange} }
\end{figure}

Using the zero mode operators $\alpha_i$, we can develop a 1+1D CFT understanding of the braiding
of the twist defects. First, consider how the quasiparticle loop operators are transformed under a
clockwise exchange of the defects $1$ and $2$, which we denote by the unitary operator
$B_{12}^\dagger$. The result can be inferred simply by the diagrams; we draw the physical processes associated with
$\alpha_1$ and $\alpha_2$, and then we consider the effect of the clockwise braid, keeping the open ends
fixed. From Fig. \ref{1dZ2exchange}, we can see that $B_{12}^\dagger \alpha_{1} B_{12} = e^{i\varphi}\alpha_{2}$
and $B_{12}^\dagger \alpha_{2} B_{12} = e^{i\theta} \alpha_{2} \alpha_{1}^\dagger \alpha_{2}$. We can fix the relative phase
$e^{i \varphi} = e^{i\theta}$ by using the fact that the braiding of $1$ and $2$ should keep invariant the
eigenvalues of the loop operator $W_1(a) = \alpha_1^\dagger \alpha_2$ that encircles the pair.
We can partially fix the remaining phase by observing that
the $Z_2$ layer exchange symmetry implies $B_{12}^\dagger \beta_1 B_{12} = e^{i\varphi} \beta_2$ and
$B_{12}^\dagger \beta_{2} B_{12} = e^{i\varphi} \beta_{2} \beta_{1}^\dagger \beta_{2}$. Using
(\ref{c2indZM}), we can then determine that $e^{i2\varphi} =  e^{-2i \pi /(m-l)}$. Thus, we find:
\begin{align}
\label{braidingZM}
B_{12}^\dagger \alpha_1 B_{12} &=  e^{i \pi k - i \pi /(m-l)}\alpha_2 ,
\nonumber \\
B_{12}^\dagger \alpha_2 B_{12} &=  e^{i \pi k -i \pi /(m-l)}\alpha_2 \alpha_1^\dagger \alpha_2,
\end{align}
where the integer $k = 0$ or $1$ indicates a remaining ambiguity in the phase $e^{i \varphi}$ that we have
not yet fixed. This implies that the physical loop operators transform as:
\begin{align}
B_{12}^\dagger W_1(a_1) B_{12} &= W_1(a_1)
\nonumber \\
B_{12}^\dagger W_1(b_1) B_{12} &= e^{i \pi k + i \pi/(m-l)} W_1^\dagger(a_1) W_1(b_1)
\end{align}
Using the fact that $W_1^{m-l}(a_1) = W_2^{m-l}(b_1) = 1$, we can use the constraint
\begin{align}
[B_{12}, W_1^{m-l}(b_1)] = 0
\end{align}
to fix $k = 1$ in the case when $m-l$ is odd.

From the above transformation, we can find the  braid matrix, $B_{12}$.
First, we pick a basis of the $|m-l|$ states:
\begin{align}
W_1(b_1) |n \rangle &= | n + 1 \text{ mod } |m-l|\rangle,
\nonumber \\
W_1(a_1) |n \rangle &= e^{2\pi i n/(m-l)} | n \rangle.
\end{align}
Then, consider
\begin{align}
\label{cftbraidZ2}
B_{12}^\dagger | n \rangle &= e^{i\theta} e^{i \pi k n + i n \pi /(m-l)} (W_1^\dagger(a_1) W_1(b_1))^n |0 \rangle
\nonumber \\
&= e^{i\theta} e^{i \pi k n -i \pi n^2/(m-l) } |n\rangle,
\end{align}
where the overall phase $e^{i\theta}$ depends on details and is not topological.
Recall that $k = 1$ when $m-l$ is odd. When $m-l$ is even, we cannot fix $k$ using this approach; the
two different choices of $k = 0$ or $1$ are related by a basis transformation
$|n \rangle \rightarrow |n + (m-l)/2 \rangle$. Alternatively, note that
$B_{12}(k = 0) = W_1^{-(m-l)/2}(b_1) B_{12}(k = 1) W_1^{(m-l)/2}(b_1)$ when $m-l$ is even.

When $m-l$ is odd, we see that the braid matrix $B_{12}$ computed in this way agrees precisely with that
obtained through the bulk geometric approach of Section \ref{bulkZ2} (see eqn. (\ref{Z2Dehn1}) ).
When $m-l$ is even, we see that the case $k = 0$ agrees with (\ref{Z2Dehn1}).

%where $W_1(a_{1}) | 0 \rangle = e^{i \delta} |0 \rangle $. The phase $\delta + \varphi$ can be fixed by
%requiring $B_{12}| |m-l| \rangle = B_{12}|0 \rangle$, so that $(\delta + \varphi) =(2n - (m-l)/2$,
%and therefore the braid matrix is
%\begin{align}
%B_{12;k} = e^{i \phi} \delta_{ab} e^{i 2\pi (q - k/2)^2/2(m-l) },
%\end{align}
%where $k$ is odd if $m-l$ is odd, and is even if $m - l$ is even. The dependence of $k$ is difficult to fix
%within this approach, which relies on only general considerations. However, observe that
%\begin{align}
%B_{12;k} = W(b)B_{12;k-2}W^\dagger(b).
%\end{align}
%Therefore, we see that the braiding, combined with the transformations $W(a)$ and $W(b)$, generate precisely
%the same group of transformations as obtained in Section ??.

\subsection{An alternative approach: braiding without moving the defects}

Now, instead of carrying out the braiding by a continuous counterclockwise motion of the twist defects,
we will consider a somewhat different protocol that can be implemented by considering
a path purely in the 1D Hamiltonian of the CFT.\cite{alicea2010b,bonderson2012} This protocol may be more readily realizable
in a physical system by, for example, gating. It also helps us compare more directly with the braiding statistics
of related defects obtained in \Ref{lindner2012,clarke2012,cheng2012, vaezi2012}.

%XLnote: I agree that the relation is indirect. We can leave out the discussion.

\begin{figure}
\centerline{
\includegraphics[width=3.6in]{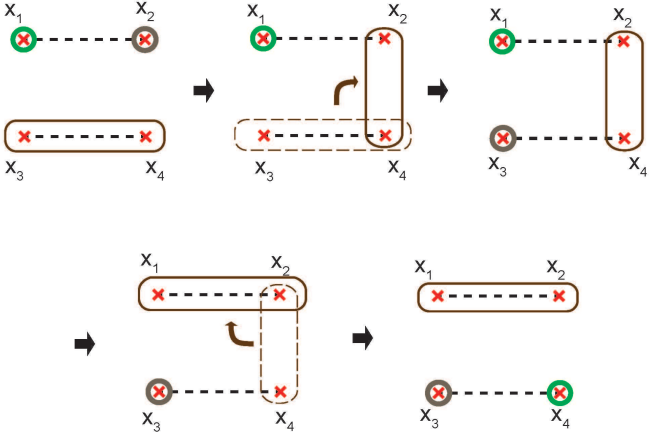}
}
\caption{
\label{1dBraidZ2}
Illustration of 1D protocol for carrying out an effective braiding process. This protocol does not require
a continuous motion in both directions and may be easier to implement in many physical realizations,
for instance through gating.
}
\end{figure}

First, let us define the following Hamiltonian:
\begin{align}
H_{ab} = -|t| (e^{-i 2\pi \theta_{ab}/(m-l)} \alpha_a^\dagger \alpha_b + H.c.),
\end{align}
which couples the zero modes located at the $a$ and $b$ defects. This has the effect of generating an energy gap in
the degenerate subspace formed by the $a$ and $b$ defects. The phases $\theta_{ab}$
determine the eigenvalue of $\alpha_a^\dagger \alpha_b$ on the ground state.

Now, we can consider the following two-step process:
\begin{eqnarray}
H(\tau) &=&\left\{\begin{array}{cc}H_{2\rightarrow 3}=(1-\tau) H_{34} + \tau H_{24},&\tau\in[0,1]\\
H_{1\rightarrow 4}=(2-\tau) H_{24} + (\tau-1) H_{12},&\tau\in[1,2]\end{array}\right.\nonumber\\
\end{eqnarray}
As is shown in Fig.  \ref{1dBraidZ2}, in the first half of the process $\tau\in [0,1]$, the Hamiltonian $H_{2 \rightarrow 3}$ moves the zero mode at defect $2$ to defect $3$. In the second half, the Hamiltonian $H_{1\rightarrow 4}$ moves the zero mode at defect $1$ to $4$.
%where we first apply $H_{1 \rightarrow 2}$, taking $u: 0 \rightarrow 1$, and subsequently we apply $H_{2 \rightarrow 3}$, taking $v: 0 \rightarrow 1$.
We set $\theta_{12} = \theta_{34}$ so that
up to a translation, this is a closed path in the Hamiltonian projected onto the low energy subspace, which exchanges the zero modes $\alpha_1$ and $\alpha_2$.
It is therefore a one-dimensional protocol for braiding the defects that does not require motion in
both directions and, depending on the realization of the twist defects, can be realized physically by gating.\cite{alicea2012review}

In order to understand the effect of these processes on the ground state subspace,
we first observe that the operators
\begin{align}
\mathcal{O}_1 &= \alpha_2 \alpha_3^\dagger \alpha_4,
\nonumber \\
\mathcal{O}_2 &= \alpha_1^\dagger \alpha_2 \alpha_4^\dagger,
\end{align}
commute with the two processes,\cite{lindner2012,clarke2012} respectively:
\begin{align}
[H_{2\rightarrow 3}, \mathcal{O}_1] &= 0,
\nonumber \\
[H_{1 \rightarrow 4}, \mathcal{O}_2] &= 0.
\end{align}
Using this, we can now obtain the effect of this process on the zero modes $\alpha_1$ and $\alpha_2$.
We let $\mathcal{P}_{a \rightarrow b} (\tau)$ be the projector onto the ground state sector of
$H_{a \rightarrow b}(\tau)$. First, we define integers $k_1$, $k_2$, and $k_3$ such that:
\begin{align}
\theta_{34} &\in (k_1 -1/2, k_1 + 1/2),
\nonumber \\
%\theta_{12} &\in (k_3 -1/2, k_3 + 1/2),
%\nonumber \\
\theta_{24} &\in -(k_2 - 1/2, k_2+1/2) - (m-l+1)/2.
\end{align}
First, we find
\begin{align}
\mathcal{P}_{2 \rightarrow 3} (0) \mathcal{O}_1 \mathcal{P}_{2 \rightarrow 3} (0) &= e^{2\pi i k_1/(m-l)} \alpha_2,
\nonumber \\
\mathcal{P}_{2 \rightarrow 3} (1) \mathcal{O}_1 \mathcal{P}_{2 \rightarrow 3} (1) &= e^{-i \pi + i \pi (2k_2 - 1)/(m-l)} \alpha_4 \alpha_3^\dagger \alpha_4.
\end{align}
Here, the projection $\mathcal{P}_{2 \rightarrow 3} (\tau) \mathcal{O}_1 \mathcal{P}_{2 \rightarrow 3} (\tau)$ keeps track of the evolution of the zero mode while we increase $\tau$ from $0$ to $1$. When $\tau=0$, i.e. when $\alpha_3$ and $\alpha_4$ are coupled, we see from the first equation that $\alpha_2$ is a zero mode of the system. When $\tau$ reaches $1$, namely when $\alpha_2$ and $\alpha_4$ are coupled, the second equation indicates that the zero mode operator evolves to $\alpha_4 \alpha_3^\dagger \alpha_4$ with an additional $U(1)$ phase. Similarly, for the second half of the process ($\tau$ increases from $1$ to $2$), we use another projection $\mathcal{P}_{1 \rightarrow 4} (\tau) \mathcal{O}_2 \mathcal{P}_{1 \rightarrow 4} (\tau)$ to follow the change of the zero mode:
\begin{align}
\mathcal{P}_{1 \rightarrow 4} (1) \mathcal{O}_2 \mathcal{P}_{1 \rightarrow 4} (1) &= e^{-i\pi + i\pi (2k_2 - 1)/(m-l)} \alpha_1^\dagger,
\nonumber \\
\mathcal{P}_{1 \rightarrow 4} (2) \mathcal{O}_2 \mathcal{P}_{1 \rightarrow 4} (2) &= e^{i2\pi k_1/(m-l)} \alpha_4^\dagger.
\end{align}
Thus, in the second half of the process, the zero mode evolves from $\alpha_1^\dagger$ to $\alpha_4^\dagger$ with a additional $U(1)$ phase factor. Also, we note that $\alpha_4 \alpha_3^\dagger \alpha_4$ commutes with the second process, $H_{2\rightarrow 3}(\tau)$, and so is unchanged.
After the process is over, the configuration of the defects are actually equivalent to the starting point and we can relabel $\alpha_4$ as $\alpha_2$ and $\alpha_3$ as $\alpha_1$. Therefore, we conclude, in terms of the braid matrix,
\begin{align}
B_{12}^\dagger \alpha_2 B_{12} &= - e^{i \frac{i\pi}{m-l} (2 k_2 - 2k_1 -1)}\alpha_2 \alpha_1^\dagger \alpha_2,
\nonumber \\
B_{12}^\dagger \alpha_1 B_{12} &= - e^{i \frac{\pi}{m-l} (2k_2 -2k_1 - 1) }\alpha_2.
\end{align}
From this, and following the same steps as in the previous section, we find that the 1D braiding protocol gives
\begin{align}
B_{12} |n \rangle = e^{i\theta} e^{i \pi (n-(k_2 - k_1))^2/(m-l)} |n \rangle.
\end{align}
Comparing with (\ref{cftbraidZ2}), we see that this 1D braiding protocol yields an extra integer degree of freedom,
$(k_2 - k_1)$, which depends on the phases $\theta_{24}$ and $\theta_{34}$. We can interpret these braidings as
the pure braiding of (\ref{cftbraidZ2}), combined with the action of the Wilson loop operators $W_1(b_1)$.

\section{$Z_2$ charge conjugation twist defects in single-component Abelian states}
\label{Z2onecomp}

\subsection{Charge conjugation twist defects}

Ref. \onlinecite{lindner2012,clarke2012,cheng2012, vaezi2012} considered another type of extrinsic defect created by the coupling between a superconductor and the edge states of a fractional quantum Hall state or a fractional quantum spin Hall state. In a fractional quantum Hall state, the defect occurs between the normal inter-edge tunneling region and the superconducting pairing region. In a fractional quantum spin Hall state, the defect occurs between superconducting and ferromagnetic regions along the edge. In both systems, the role of the superconductor is to break charge conservation. In the following we will not review their proposals in more detail, but we will discuss the understanding of this defect as a twist defect, and its relation to the $Z_2$ twist defect discussed in the previous section.

To begin with, we consider Abelian topological states that are described by $U(1)_N$ CS theory, namely the
$1/N$-Laughlin FQH states:
\begin{align}
\mathcal{L} = \frac{N}{4\pi} \epsilon^{\mu \nu \lambda} a_\mu \partial_\nu a_\lambda.
\end{align}
These states have $N$ quasiparticles, $\{ \gamma_i \}$ for
$i = 1, \cdots, N$, with topological spin $\theta_a$ and fractional charge $q_a$
\begin{align}
\theta_a = \frac{\pi a^2}{N}, \;\; q_a = a/N.
\end{align}
$\theta_a$ is defined modulo $\pi$ when $N$ is odd (fermions) and modulo $2\pi$ when $N$ is even (bosons),
while $q_a$ is defined modulo 1. The mutual statistics between quasiparticles $\gamma_a$ and $\gamma_b$ is given by
\begin{align}
\theta_{ab} = 2\pi a b/N,
\end{align}
which is defined modulo $2\pi$.

We see that when charge conservation is broken, this theory has a $Z_2$ symmetry associated with
\begin{align}
\gamma_a \rightarrow \gamma_{N-a}.
\end{align}
In the CS field theory, this symmetry is implemented as
\begin{align}
a \rightarrow - a.
\end{align}

This implies that we can consider twist defects associated with this $Z_2$ symmetry. As in the previous section,
we align all the twist defects, and we cut the system along a line to obtain counterpropagating
edge states. The electron operator on each edge is:
\begin{align}
\Psi_{eL} = e^{i N \phi_L}, \;\; \Psi_{eR} = e^{-i N \phi_R}.
\end{align}
Now, in the untwisted $A$ regions, we consider the usual hopping,
$\Psi_{eL}^\dagger \Psi_{eR} + H.c. \propto \cos(N \phi)$. In the twisted regions,
we apply the $Z_2$ action to one of the chiral edge states: $\phi_R \rightarrow -\phi_R$,
to get the tunneling term $\Psi_{eL}^\dagger \Psi_{eR}^\dagger + H.c. \propto \cos(N \theta)$,
where
\begin{align}
\phi &= \phi_L + \phi_R,
\nonumber \\
\theta &= \phi_L - \phi_R.
\end{align}
Thus we have:
\begin{align}
\label{scfmFTI}
\delta H_t &= \frac{g}{2} \left\{ \begin{array}{ccc}
(\Psi_\uparrow \Psi_\downarrow + H.c.) & \text{ if } & x \in A_i \\
(\Psi_\uparrow^\dagger \Psi_\downarrow + H.c.)  & \text{ if } & x \in B_i \\
\end{array} \right.
\nonumber \\
&= g \left\{ \begin{array}{ccc}
\cos (N\theta) & \text{ if } & x \in A_i \\
\cos (N\phi)& \text{ if } & x \in B_i \\
\end{array} \right.
\end{align}
This is precisely what has been considered in Ref. \onlinecite{fu2009d,lindner2012,clarke2012,cheng2012}, where it was
observed that the defects have quantum dimension $\sqrt{2N}$.

\subsection{Relation to genons in two-component states}

Now let us consider the $Z_2$ genons in the two-component theories with $K$-matrix
$K = \left( \begin{matrix} m & l \\ l & m \end{matrix} \right)$. In this theory, the two kinds of tunneling
terms that we add are (see eqn. (\ref{2compTunn}))
\begin{align}
\delta \mathcal{H}_{t} =
g \cos\left( \frac{m+l}{2} \phi_+\right)  \left\{ \begin{array}{ccc}
\cos( \frac{m-l}{2} \phi_-) & \text{ if } & x \in A_i \\
\cos( \frac{m-l}{2} \tilde{\phi}_-) & \text{ if }& x \in B_i\\
\end{array} \right.
\end{align}
Now observe that if we define
\begin{align}
\phi_{L\pm} &= \phi_{L1} \pm \phi_{L2},
\nonumber \\
\phi_{R \pm} &= \phi_{R1} \pm \phi_{R2},
\end{align}
and further define
\begin{align}
\theta_{-} = \phi_{L-} - \phi_{R-} = \tilde{\phi}_-,
\end{align}
then
\begin{align}
\label{2compTun}
\delta \mathcal{H}_{t} =
g \cos\left( \frac{m+l}{2} \phi_+\right)  \left\{ \begin{array}{ccc}
\cos( \frac{m-l}{2} \phi_-) & \text{ if } & x \in A_i \\
\cos( \frac{m-l}{2} \theta_-) & \text{ if }& x \in B_i\\
\end{array} \right.
\end{align}
Since both tunneling terms contain $\cos( \frac{m+l}{2} \phi_+)$, $\phi_+$ is pinned everywhere and in the
low energy sector can be replaced by a constant. Now we see that in the Luttinger liquid theory, the
two different tunneling terms are identical to the superconductor - ferromagnetic terms in the FQSH edge (see (\ref{scfmFTI})).
This explains the agreement between the braiding matrices computed from the Dehn twists of the high genus
surface (Section \ref{Z2twocomp}), and that computed in Ref. \onlinecite{lindner2012,clarke2012,cheng2012}.

In (\ref{scfmFTI}), the defects have quantum dimension $d = \sqrt{2N}$, while in the two-component case,
they have quantum dimension $d = \sqrt{|m-l|}$. Therefore, $d^2$ is not restricted to be even in the
latter case. The reason is that in the two-component case, the coefficient of the boson fields in the
$\cos$ terms can have half-integer values (see \ref{2compTun}), while it is restricted to be integer in the
FQSH setup (\ref{scfmFTI}). This can be traced to the fact that the boson fields $\theta_-$ and $\phi_-$
are compactified on a circle: $(\theta_-, \phi_-) \sim (\theta_- + 2\pi, \phi_- + 2\pi)$. Therefore,
$\cos( \frac{m-l}{2} \theta_-)$ and $\cos( \frac{m-l}{2} \phi_-)$ are not individually invariant under such a gauge transformation
if $m-l$ is odd. However, in the two-component case, this is allowed, because whenever
one of $\phi_{LI}$ or $\phi_{RI}$ are advanced by $2\pi$, both $\theta_-$ and $\phi_+$ or $\phi_-$ and $\phi_+$ will
change by $2\pi$, so that the products $\cos( \frac{m+l}{2} \phi_+) \cos( \frac{m-l}{2} \theta_-)$
and $\cos( \frac{m+l}{2} \phi_+) \cos( \frac{m-l}{2} \phi_-)$ are invariant.

In light of this, we note that even in the $n$-layer FQH states, one can introduce superconductivity in addition
to twisted tunnelings in order to create zero modes. We expect then that one can always map such situations
onto a $2n$-layer system without superconductivity but with twisted tunnelings.

\section{$Z_2$ genons in two-component non-Abelian states and universal quantum computing}
\label{Z2na}

A large part of the analysis in the last section can be generalized to non-Abelian states.
In this section we briefly comment on the somewhat more exotic possibility of twist defects in non-abelian states.
Let $G$ denote any non-abelian topological state, and let us consider two independent copies of such a state,
which we will label $G \times G$. For example, we may take $G$ to be an Ising topologically ordered state.
The $G \times G$ state has a $Z_2$ symmetry in its topological quantum numbers associated with exchanging the two copies.
$n$ pairs of genons on a sphere will therefore lead to a single copy of $G$, on a genus $g = n-1$ surface. The
ground state degeneracy $S_g$ is then given by the general formula
\begin{align}
S_g = D^{2(g - 1)} \sum_{i=1}^{N_{qp}} d_i^{-2(g-1)},
\end{align}
where $d_i$ is the quantum dimension of the $i$th quasiparticle, $N_{qp}$ is the number of quasiparticles in $G$,
and $D =\sqrt{ \sum_{i=1}^{N_{qp}} d_i^2}$ is the total quantum dimension of $G$. In the limit of large $g$, we see
that $S_g \sim D^{2n}$, which shows that  the genons have a quantum dimension
\begin{align}
d_{Z_2} = D.
\end{align}

In the case of genons in the Ising $\times$ Ising theory, we see that $d_{Z_2} = 2$,
and can effectively yield a single Ising theory on a high genus surface. The braiding of the twist
defects corresponds to Dehn twists of the Ising theory on the high genus surface.
\begin{figure*}
\centerline{
\includegraphics[width=6.5in]{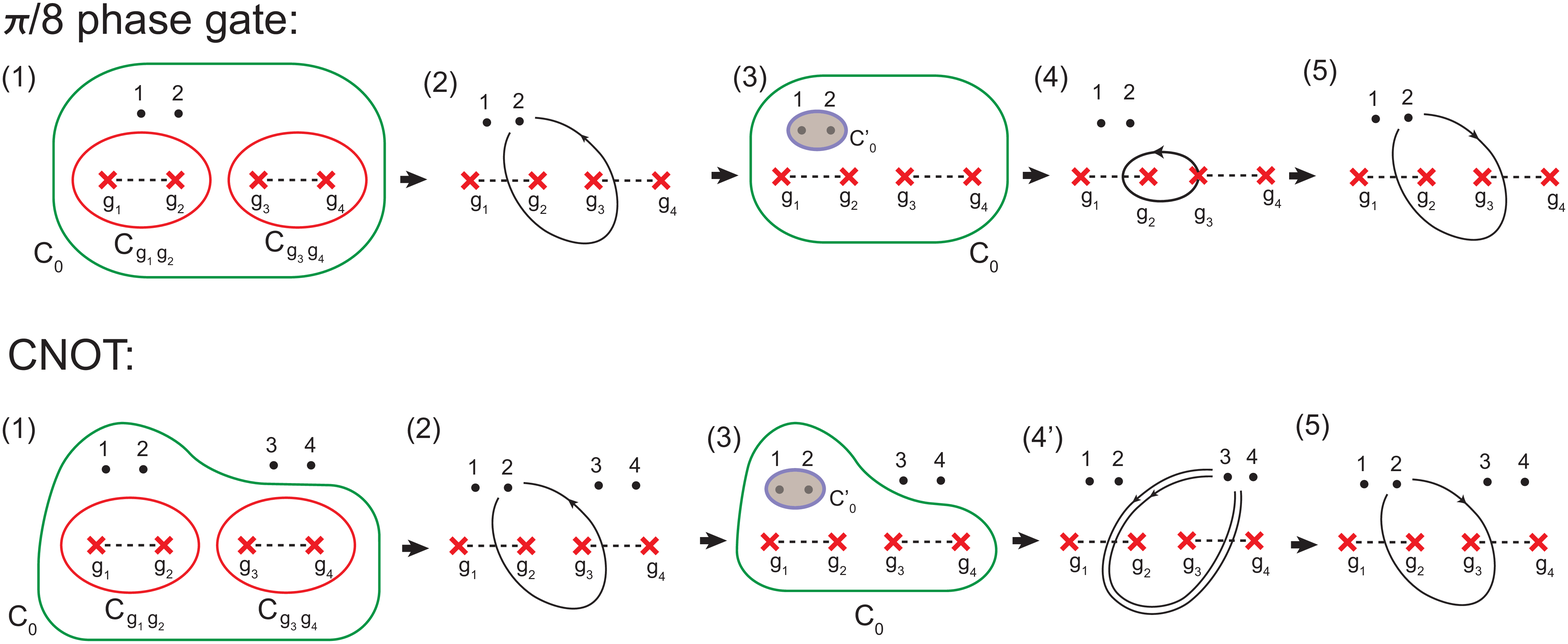}
}
\caption{Illustration of operations (1), (2), (3), (4) and (4'), described in the main text for performing the CNOT and $\pi/8$ phase gates. The black dots are $(\sigma,1)$ topological quasiparticles. The red circles indicate that the topological charge of the two twist defects enclosed in each  circle is trivial. The black loops with arrows stand for quasi-particle paths. The shaded region enclosed by a blue circle $C_0'$ stands for a measurement which projects the state in loop $C_0'$ to the trivial sector $(1,1)$.
\label{universalTQC} }
\end{figure*}
While the Ising theory by itself is known to be non-universal for TQC, it is known that
it can be made universal if it were possible to add handles in the space and carry out
Dehn twists\cite{bravyi200universal,freedman2006universal}. The $Z_2$ genons
provide a physical realization of such a possibility and therefore can be used to render universal
for TQC a state that without the twist defects is non-universal for TQC.

Here we will briefly describe how to use the genons to implement the necessary computational
gates in the Ising $\times$ Ising state, adapting the basic ideas of
Ref. \onlinecite{bravyi200universal,freedman2006universal}. For more comprehensive reviews of the Ising topological state, we refer the readers to books and review articles\cite{nayak2008,zhwang2010}.
The Ising $\times$ Ising state contains 9 distinct
quasiparticles, of the form $a \times b$, where $a,b = 1, \psi, \sigma$ are the three quasiparticles
of the Ising state, which have spins $h_1 = 0$, $h_\psi = 1/2$ and $h_\sigma = 1/16$.
In the presence of genons, the non-contractible loops surrounding the genons
can be mapped to non-contractible loops of a single Ising theory on a high genus surface; therefore
each non-contractible loop surrounding the genons can be labelled by one of three topological
charges: $1$, $\psi$, or $\sigma$.

A universal set of gates for quantum computing is obtained in terms of the single-qubit '$\pi/8$' phase
gate:\cite{bravyi200universal,freedman2006universal}
\begin{align}
G_1 = \left( \begin{matrix} 1 & 0 \\ 0 & e^{2\pi i /8} \end{matrix} \right),
\end{align}
and two two-qubit gates:
\begin{align}
G_2 = \left(\begin{matrix}
1 & 0 & 0 & 0 \\
0 & 1 & 0 & 0 \\
0 & 0 & 1 & 0 \\
0 & 0 & 0 & -1 \end{matrix} \right),
\;\; G_3 = \frac{1}{\sqrt{2}} \left( \begin{matrix}
1 & 0 & 0 & -i \\
0 & 1 & -i & 0 \\
0 & -i & 1 & 0\\
-i & 0 & 0 & 1 \\
\end{matrix} \right).
\end{align}
$G_2$ is, after a basis transformation, the controlled-not (CNOT) gate.\cite{freedman2006universal}
Each pair of $\sigma$ quasiparticles in the Ising state
can fuse to either $1$ or $\psi$, and thus these two possible states form the two states of a qubit.
Therefore $n$ qubits requires $2n$ $\sigma$ particles. While $G_3$ can be implemented by 
braiding the quasiparticles in the Ising theory, $G_1$ and $G_2$ cannot be implemented by 
braiding quasiparticles alone. However they can be implemented by creating genons and braiding them around the
quasiparticles.

%\begin{figure}
%\centerline{
%\includegraphics[width=4in]{tqc2-01.eps}
%}
%\caption{Illustration of operations (1), (2), (3), (4) and (4'), described in the main text for performing the CNOT and %$\pi/8$ phase gates. The red circles indicate that the topological charge of the two twist defects enclosed in each  circle %is trivial. The black loops with arrows stand for quasi-particle paths.
%\label{universalTQC} }
%\end{figure}

In order to implement $G_1$, we consider two quasiparticles of type $\sigma \times 1$,
which we label $1$ and $2$. The pair $(1,2)$ fuses to the channel $x \times 1$, where $x$ can be $1$ or $\psi$.
The two channels form the two states of a qubit that can be measured by interferometry around a loop surrounding
the two quasiparticles, which we denote as $C_0$.

We implement the following processes (see Fig. \ref{universalTQC}):

(1) Create two pairs of genons out of the vacuum inside the loop $C_0$, which we label $g_1, \ldots, g_4$. Let $C_{g_i, g_{i+1}}$ label the loop
surrounding $g_i$ and $g_{i+1}$. After creating the genons, ensure that the topological charges around the loops $C_{g_1, g_2}$ and
$C_{g_3,g_4}$ are trivial, which can be done by performing a measurement.

(2) Braid quasiparticle $2$ around $C_{g_2, g_3}$.

(3) Check with an interferometry measurement that the charge around $C_0'$ is $1$. If not, then re-annihilate the genons and restart from step (1) until we find the unit charge around $C_0'$.

(4) Perform a double-exchange of $g_2$ and $g_3$.

(5) Undo step (2) by taking quasiparticle 2 around $C_{g_2 g_3}$ in the opposite direction as compared with step (2), and
then annihilate the genons.

\begin{figure}
\centerline{
\includegraphics[width=2in]{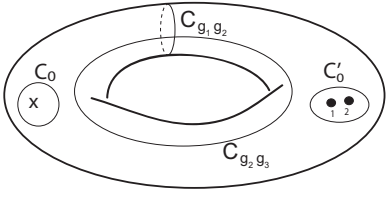}
}
\caption{Mapping Fig. \ref{universalTQC} to single Ising theory on a torus. The loops $C_0$, $C_0'$, $C_{g_1 g_2}$, and $C_{g_2g_3}$ are depicted.
\label{torusQbit} }
\end{figure}

This procedure implements $G_1$. In step (1), we ensure that the loops $C_{g_1, g_2}$ and $C_{g_3,g_4}$ carry trivial charges 
so that the charge of $C_0$ is unchanged before and after the genon-pair creation. By the high genus surface mapping, 
the region inside the loop $C_0$ can be understood as a single layer of the Ising theory on a torus with a puncture, 
denoted by $C_0$, and two quasiparticles on it (see Fig. \ref{torusQbit}). We will denote the topological charge through a loop $C$,
in this mapping to a single Ising theory on a torus, as $W(C)$. Therefore $W(C)$ will in general be some superposition 
of $1$, $\psi$, and $\sigma$. The charge of the puncture is $W(C_0) = x$, the state 
of qubit. Let us consider the topological charge through $C_0'$ and $C_{g_2 g_3}$, which we will denote as
$W(C_0') \otimes W(C_{g_2g_3})$. After step (1), this is $x \otimes \frac{1}{2}(1 + \psi + \sqrt{2}\sigma)$. 
This follows from the fact that $W(C_{g_1 g_2}) = 1$, combined with the properties of the modular $S$-matrix:
\begin{align}
S = \frac{1}{2}\left( \begin{matrix} 
1 & \sqrt{2} & 1 \\
\sqrt{2} & 0 & -\sqrt{2} \\
1 & -\sqrt{2} &  1 \\
\end{matrix} \right),
\end{align}
which determines $W(C_{g_2 g_3})$ in terms of $W(C_{g_1 g_2})$. After step (2), $W(C_{g_1 g_2})$ changes
to $\sigma$, while $W(C_0') \otimes W(C_{g_2 g_3}) \rightarrow x  \otimes \frac12(1 - \psi) + \frac{1}{\sqrt{2}} (\psi x ) \otimes \sigma$. 
This is simply because the fusion channel of 2 $\sigma$ particles is flipped when one of them is braided around a 
third $\sigma$ particle, which in this case is the $\sigma$ state of $W(C_{g_2 g_3})$. 
Therefore, step (3), which projects onto the sector where $W(C_0') = 1$, projects
$W(C_{g_2 g_3})$ to be $\frac{1}{\sqrt{2}}(1 - \psi)$ if $x = 1$, or $\sigma$ if $x = \psi$. 
%After step (2), the topological charge of $C_{g_1, g_2}$, which is one of the non-contractible loops on the 
%torus, is $\sigma$. Using the modular $S$-matrix we know that the topological charge of $C_{g_2,g_3}$ is in a 
%superposition state $\left(1-\psi\right)/\sqrt{2}$ and $\sigma$. This state is entangled with the state of the 
%particle pair $1,2$, since the state of $1,2$ will remain invariant if the charge of $C_{g_2,g_3}$ is $\left(1-\psi\right)/\sqrt{2}$, but it will flip between $1$ and $\psi$ if the latter is $\sigma$. 
Another way to see the above result is as follows. Since step (3) projects onto the sector where 
$W(C_0') = 1$, we can fill in the interior of $C_0'$ with vacuum, leaving us with a
torus with a single puncture, denoted by the loop $C_0$, which encodes the state of the qubit. 
If the initial state of the $1,2$ pair is $x = 1$, then the fact that all punctures of the torus are filled in with vacuum, 
and $W(C_{g_1 g_2}) = \sigma$, together with $S$, implies that $W(C_{g_2 g_3})$ is $\left(1-\psi\right)/\sqrt{2}$.
%then $C_{g_2,g_3}$ is projected to $\left(1-\psi\right)/\sqrt{2}$. 
In contrast, if the initial state is $x=\psi$, we are left with a torus with a single puncture with charge $W(C_0) = \psi$ and
with $W(C_{g_1 g_2}) = \sigma$. Using the fact that the $S$-matrix in the presence of such a puncture
is $S^\psi_{\sigma i} = \delta_{i \sigma}$ \cite{bravyi200universal,freedman2006universal} for $i  =1$, $\sigma$, or $\psi$,
then it follows that $W(C_{g_2,g_3}) = \sigma$. 
%it can be shown that, due to the properties of the modular $S$-matrix, the topological charge of $C_{g_2, g_3}$ is $(1 - \psi)/\sqrt{2}$.
In step (4), the double exchange, $g_2$ and $g_3$, which implements the
double Dehn twist in the geometrical picture, has an eigenvalue of $1$ for the state $W(C_{g_2 g_3}) = \left(1-\psi\right)/\sqrt{2}$ , 
because $e^{2(2\pi i h_1)} = e^{2(2\pi i h_\psi)} = 1$, and $(e^{2 \pi i/16} )^2 = e^{2\pi i /8}$ for the state $W(C_{g_2 g_3}) = \sigma$.
%On the other hand, if $x=\psi$, then the topological charge of $C_{g_2, g_3}$ is $\sigma$.\cite{freedman2006universal} In this case, the double exchange of the genons gives a phase corresponding to twice the spin of $\sigma$: $(e^{2 \pi i/16} )^2 = e^{2\pi i /8}$.
In step (5), undoing step (2)
%reverts the topological charge of $C_{g_1 g_2}$ and $C_{g_3 g_4}$ back to the identity,
restores the original state of the $1,2$ pair and unentangles it with the state of the genons, so that annihilating the genons afterwards does not affect the qubit state.
%XLnote1202: three points: 1. There is no step 5 in the figure. 2. Please see if my description above is correct. 3. I think the more important role of step 5 is to restore the state of 1,2 rather than that of g1g2 and g3g4. I am not sure if the state of g1g2 and g3g4 are restored. Are they?

In order to implement $G_2$, consider two pairs of quasiparticles, all of type $\sigma \times 1$, which we label
$1,2,3,4$. Then, perform the same steps (1), (2), (3) and (5) as above, replacing (4) with (4'):

(4') Braid the pair $(3,4)$ around the loop $C_{g_2, g_3}$.

Based on the previous discussion, the step (4') gives a $+1$, unless $(3,4)$ is in the state $\psi \times 1$
\it and \rm $W(C_{g_2,g_3}) = \sigma$ (\it ie \rm if $W(C_0) = \psi$), in which case step (4') gives $-1$.

\section{$Z_3$ twist defects and braiding in three-component Abelian FQH states}
\label{Z3threecomp}

To understand more generic behavior of twist defects, here we extend the analysis of quantum dimensions and braiding of twist defects to $Z_3$ twist defects
in three-component Abelian states, which are described by $U(1) \times U(1) \times U(1)$ CS theory:
\begin{align}
\mathcal{L} = \frac{K_{IJ}}{4\pi} a^I \partial a^J,
\end{align}
for a rank-3 $K$-matrix. Here the $Z_3$ nature of the twist defects we consider is associated with the $Z_3$ cyclic permutation
symmetry among the three components. Such $Z_3$ defects can be realized by the lattice dislocations in the
Chern number $3$ topological nematic states\cite{barkeshli2012prx}. In principle it can also be realized in triple layer quantum Hall states, which have also been realized experimentally \cite{Gusev2009,Jo1992}. %{\bf (cite refs).}%XLnote: Ref on 3-layer FQH is still needed.
Imposing a $Z_3$ layer symmetry implies that the $K$-matrix
depends on two integers:
%Here we extend the analysis of quantum dimensions and braiding of twist defects to $Z_3$ twist defects
%in a three-component Abelian state, described by a rank-3 $K$-matrix. We note that triple layer quantum Hall states
%have also been realized experimentally (cite refs). Imposing a $Z_3$ layer symmetry implies that the $K$-matrix
%depends on two integers:
\begin{align}
\label{KZ3}
K = \left( \begin{matrix} m & l & l \\ l & m & l \\ l & l & m \end{matrix} \right).
\end{align}
%As we will see, this case requires a more general analysis with features that do not show up
%in the case of the simpler $Z_2$ twist defects. Our analysis will rely on the 1+1D edge
%CFT picture, as the bulk geometric picture and mapping to higher genus surface
%is less convenient and harder to visualize in this case.

As we will see, this case requires a more general analysis with features that do not show up
in the case of the simpler $Z_2$ twist defects. Since the bulk geometric picture and mapping to higher-genus surface
is harder to visualize than the $Z_2$ case, we will first carry out our analysis based on the 1+1D edge
CFT picture, and then present the bulk geometric picture.

\subsection{1+1D edge CFT picture of the $Z_3$ defects}
\label{Z3CFT}

\begin{figure}
\centerline{
\includegraphics[width=3.6in]{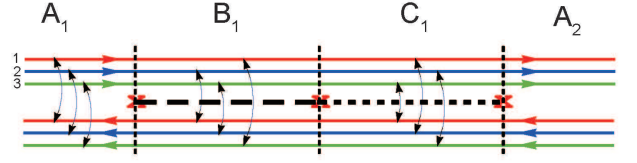}
}
\caption{Oriented all of the defects along a line and cutting yields three counterpropagating edge modes.
Since the defects are $Z_3$, there are now three different regions, $A_i$, $B_i$, and $C_i$, where we introduce
different twisted tunneling terms.
\label{C3edge} }
\end{figure}

In this case, the $Z_3$ twist defects come in groups of three. As before, we align them on a torus and cut the
system along the line to obtain counterpropagating edge states. Similar to the $Z_2$ case, the actions of the edge theory on the two sides of the cut are given by (\ref{edgeAc}) and (\ref{RedgeAc}) respectively. As shown before, we can obtain the commutator (\ref{1dCommutatorL}) and (\ref{1dCommutatorR}) by quantizing the theory. However, instead of using (\ref{1dZ2Klein}), here we choose a different scheme to take care of the correct statistics between electrons on different layers. We write the electron operators as:
\begin{align}
\Psi_{eLI} &=\eta_{LI} e^{iK_{IJ}\phi_{LI}}, \nonumber \\
\Psi_{eRI} &=\eta_{RI} e^{-iK_{IJ}\phi_{RI}},
\end{align}
where $\eta_{LI}$ and $\eta_{RI}$ are Klein factors that satisfy
\begin{align}
&~~~~~~~~~~~~~\eta_{LI}^2=\eta_{RI}^2=1,~\eta_{LI}^\dag=\eta_{LI},~\eta_{RI}^\dag=\eta_{RI}, \nonumber \\
&\left\{ \begin{array}{cc}
\left[\eta_{LI},\eta_{LJ}\right]=\left[\eta_{RI},\eta_{RJ}\right]=\left[\eta_{LI},\eta_{RJ}\right]=0, & \text{for } m-l \text{~even} \\
\{\eta_{LI},\eta_{LJ}\}=\{\eta_{RI},\eta_{RJ}\}=\{\eta_{LI},\eta_{RJ}\}=0, & \text{for } m-l \text{~odd}. \\
\end{array} \right.
\end{align}
When we glue the edges back together, we now have three different regions where we introduce
different tunneling operators (see Fig. \ref{C3edge}):
\begin{align}
\delta H_{tun}= g \sum_I \left\{ \begin{array}{ccc}
\mathcal{K}_I\cos (K_{IJ} \phi_J) & \text{if } & x \in A_i \\
\mathcal{K}'_I\cos(K_{IJ} \phi_J') & \text{if } & x \in B_i \\
\mathcal{K}''_I\cos (K_{IJ} \phi_J'') & \text{if } & x \in C_i
\end{array} \right.  \label{1dZ3tunnel}
\end{align}
where now
\begin{align}
&\phi_I' = \phi_{LI} + \phi_{R (I+1) \% 3}, \;\; \phi_I'' = \phi_{LI} + \phi_{R (I+2)\%3},\nonumber \\
&\mathcal{K}_I=i^{m-l}\eta_{LI}\eta_{RI},~\mathcal{K}'_I=i^{m-l}\eta_{LI}\eta_{R(I+1)\%3}, \nonumber \\
&\mathcal{K}''_I=i^{m-l}\eta_{LI}\eta_{R(I+2)\%3}.
%\phi_1' &= \phi_{L1} + \phi_{R2}, \;\; &\phi_1'' = \phi_{L1 } + \phi_{R3},
%\nonumber \\
%\phi_{2}' &= \phi_{L2} + \phi_{R3}, \;\; &\phi_2'' = \phi_{L2} + \phi_{R1},
%\nonumber \\
%\phi_{3}' &= \phi_{L3} + \phi_{R1}, \;\; &\phi_3'' = \phi_{L3} + \phi_{R2}.
\end{align}
%\begin{figure}
%\centerline{
%\includegraphics[width=3in]{zeroModesC3-01.eps}
%}
%\caption{Physical meaning of the zero mode operators $\alpha_i$, $\beta_i$ and $\gamma_i$ for the $Z_3$ twist defect case. In %subsequent figures, we will use the bottom figure, which is more precisely described by the top figure.
%\label{ZeroModesC3} }
%\end{figure}
The factor $i^{m-l}$ helps keep the hermiticity of the operators $\mathcal{K}_I$, $\mathcal{K}'_I$ and $\mathcal{K}''_I$.
In the absence of the twist defects, we only consider a single $A$ region, and the minima of the cosine potential give
$|\text{Det } K| = |(m+2l) (m-l)^2|$ different states. These states can be labelled by integer vectors $\vec{p} = (p_+, p_{1,-2,1}, p_{1,0,-1})$, such that
\begin{align}
e^{i \phi_+} |\vec{p} \rangle &= e^{\frac{2\pi i p_+}{m+2l}} |\vec{p} \rangle,
\nonumber \\
e^{i \phi_{(1,-2,1)}} |\vec{p}\rangle &= e^{\frac{i 2\pi p_{(1,-2,1)}}{m - l }} |\vec{p} \rangle,
\nonumber \\
e^{i \phi_{(1,0,-1)}} |\vec{p}\rangle &= e^{\frac{i 2\pi p_{(1,0,-1)}}{m - l }} |\vec{p} \rangle,
\label{1dZ3DegLabel}
\end{align}
where we have defined $\phi_{(a,b,c)} = a \phi_1 + b \phi_2 + c \phi_3$, and $\phi_+ = \phi_{(1,1,1)}$.

Now consider the system with $n > 0$ triplets of twist defects. While formally the tunneling Hamiltonian in the
different regions differ only by a cyclic permutation of the layers,
there is an important difference between the regions. Since all physical operators can only be built out of electron
operators, there is a gauge symmetry in the values of $\phi_{LI}$ and $\phi_{RI}$. The following transformation:
\begin{align}
\phi_{LI} &\rightarrow \phi_{LI} + 2\pi K^{-1}_{IJ} n_J,
\nonumber \\
\phi_{RI} &\rightarrow \phi_{RI} - 2\pi K^{-1}_{IJ} n_J,
\end{align}
with $n_1$, $n_2$, $n_3 \in \mathbb{Z}$, preserves all physical operators and are thus considered gauge symmetries of the theory.
All physical operators must be gauge invariant. In the $A$ regions, the quasiparticle tunneling operators $e^{i \phi_I(x_{A_i})}$ are all gauge-invariant.
In contrast, in the $B$ and $C$ regions, $e^{i \phi'_I}$ and $e^{i\phi''_I}$ are unphysical. However, the operators
$e^{i \phi'_+ (x_{B_i})} = e^{i \phi_+(x_{B_i})}$, $e^{i \phi''_+ (x_{C_i})} = e^{i \phi_+(x_{C_i})}$, $e^{i \phi'_{(1,0,-1)}(x_{B_i}) + \phi''_{(-1,1,0)} (x_{C_i})}$
and $e^{\phi'_{(-1,1,0)}(x_{B_i})+\phi''_{(0,-1,1)}(x_{C_i})}$ are also physical, gauge-invariant operators.
The cosine potentials pin the eigenvalues of these operators. Due to the commutation relations:
\begin{align}
[\phi_I(x),\phi'_J(y)] &=i\pi(K^{-1}_{I,(J+1)\%3}-K^{-1}_{I,J})\sgn(x-y),  \nonumber \\
[\phi_I(x),\phi''_J(y)] &=i\pi(K^{-1}_{I,(J+2)\%3}-K^{-1}_{I,J})\sgn(x-y), \nonumber \\
[\phi'_I(x),\phi''_J(y)] & =i\pi(K^{-1}_{I,(J+1)\%3}-K^{-1}_{I,J})\sgn(x-y),
\end{align}
we see that the largest set of independent commuting physical operators is $e^{i \phi_I(x_{A_i})}$, $e^{i \phi_+(x_{B_i})}$,
$e^{i \phi_+(x_{C_i})}$, and $e^{i \phi'_{(1,0,-1)}(x_{B_i}) + i\phi''_{(-1,1,0)} (x_{C_i})}$.

Minimizing the cosine potential in the $A$ regions, we get $|(m+2l)(m-l)^2|$ states for each $A$ region, as in the case with no
twist defects. In the $B$ and $C$ regions, the operator $e^{i \phi_+}$ is pinned to one of $(m+2l)$ values in each region. Finally,
for each triplet of defects, the operator $e^{i \phi'_{(1,0,-1)}(x_{B_i}) + \phi''_{(-1,1,0)} (x_{C_i})}$ is pinned to one of $|m -l|$ values.
In Appendix \ref{1dZ3GSDapp}, we will provide a more detailed treatment which explicitly includes the Klein factors.

The above counting gives a total of $|(m+2l)^{3n} (m-l)^{3n}|$ states. However, there are additional constraints:\cite{barkeshli2012prx} The electric charge is a local
observable, and therefore states with different charges at the defects are not topologically degenerate. This gives
$3n-1$ constraints on each of the eigenvalues of $e^{i\phi_+}$ in the different regions. The total number of states is therefore
$|(m+2l)(m-l)^{3n}|$. Each $Z_3$ twist defect therefore has quantum dimension
\begin{align}
d_{Z_3} = |m -l |.
\end{align}
Writing the ground state degeneracy as $|\text{Det } K| |m-l|^{3n-2}$, the factor $|\text{Det } K|$ can be understood from the fact
that we started on a torus, while the factor $|m - l |^{3n-2}$ can be understood as due to the appearance of a non-trivial quasiparticle
loop algebra, which forms $3n-2$ copies of a magnetic algebra. The degeneracy $|m-l|^{3n-2}$ is equivalent to what would be obtained
in a $U(1)_{m-l}$ CS theory on a genus $g = 3n-2$ surface (the explicit mapping to the high genus surface will be explained
in Sec. \ref{Z3geoSec}). Below, we calculate the braiding of these $Z_3$ genons, which can also be related to Dehn twists of
$U(1)_{m-l}$ CS theory.

\begin{figure}
\centerline{
\includegraphics[width=2.8in]{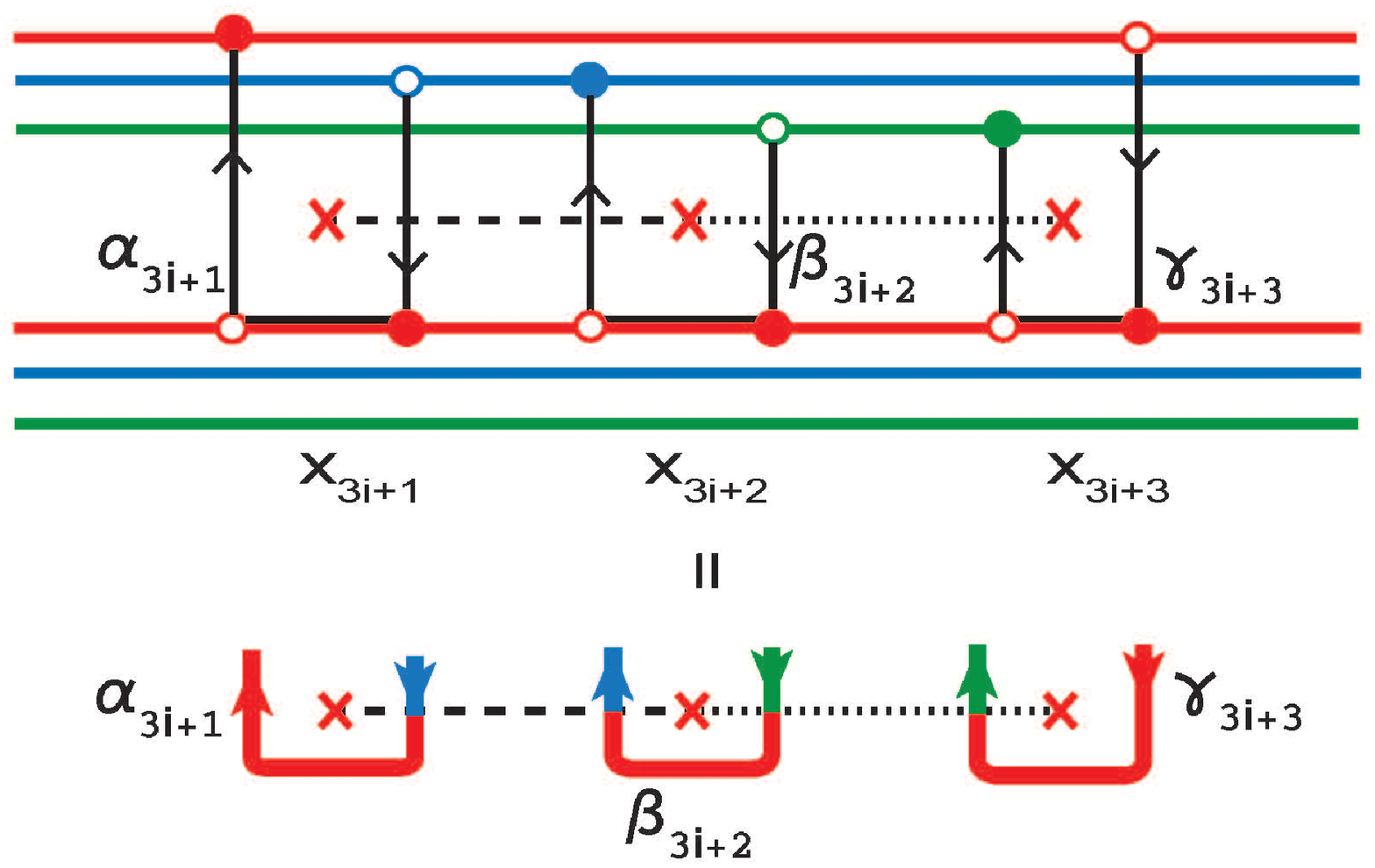}
}
\caption{Physical meaning of the zero mode operators $\alpha_i$, $\beta_i$ and $\gamma_i$ for the $Z_3$ twist defect case. In subsequent figures, we will use the bottom figure, which is more precisely described by the top figure.
\label{ZeroModesC3} }
\end{figure}

To compute the braiding, as before, we can construct zero modes localized to each domain wall, which correspond
to quasiparticle tunneling around the defects:
%Classically, within each region we have $|\text{Det } K| = |(m+2l) (m-l)^2|$ ground states. Naively, for $n$ triplets of %dislocations, this would give
%$|(m+2l)(m-l)^2|^{3n}$ ground states (The factors $\mathcal{K}_I$, $\mathcal{K}'_I$ or $\mathcal{K}''_I$ are projected to be %$1$ on the ground states in both classical or quantum theory). However the charge sector, $\phi_+ \equiv \sum_{I=1}^3 %\phi_I$, is globally pinned, so we obtain
%$|(m+2l)(m-l)^{6n}|$ ground states in the classical theory. We can pick the classical coordinates to be
%$\phi_+ = \frac{2\pi n_+}{m + 2l}$, $\phi_{(1,-2,1)}(x_i) = \phi_1 - 2 \phi_2 + \phi_3 = \frac{2\pi n_{1,2,1}}{m-l}$, and
%$\phi_{1,0,-1}(x_i) = \frac{2\pi n_{1,0,-1}}{m-l}$.
%where $i=1,2,...,n$, $x_{3i+1}\in A_i$, $x_{3i+2}\in B_i$ and $x_{3i+3}\in C_i$. $\phi_+$ can take $m+2l$ different values, %while each of the other coordinates can have $m-l$ values. Therefore, we obtain $|(m+2l) (m-l)|^{3n}$ ground states, which %shows that each $Z_3$ twist defect has quantum
%dimension $|m-l|$. As before, we can construct zero modes localized to each domain wall, which correspond
%to quasiparticle tunneling around the defects:
\begin{align}
\alpha_i = \left \{ \begin{array}{cc}
\eta_{1R}\eta_{2R} e^{i \phi_1 (x_{A_i})} e^{-i \phi_1'(x_{B_i})}, & i \text{ mod } 3 = 1\\
\eta_{1R}\eta_{2R} e^{i \phi_3'(x_{B_i})} e^{-i \phi_3''(x_{C_i})}, & i \text{ mod } 3 = 2\\
\eta_{1R}\eta_{2R} e^{i\phi_2''(x_{C_i})} e^{-i \phi_2(x_{A_{i+1}})}, & i \text{ mod } 3 = 0\\
\end{array} \right.
\end{align}
Similarly, we define $\beta_i$ :
\begin{align}
\beta_i = \left \{ \begin{array}{cc}
\eta_{2R}\eta_{3R} e^{i \phi_2  (x_{A_i})} e^{-i \phi_2' (x_{B_i})}, & i \text{ mod } 3 = 1\\
\eta_{2R}\eta_{3R} e^{i \phi_1' (x_{B_i})} e^{-i \phi_1''(x_{C_i})}, & i \text{ mod } 3 = 2\\
\eta_{2R}\eta_{3R} e^{i \phi_3''(x_{C_i})} e^{-i \phi_3  (x_{A_{i+1}})}, & i \text{ mod } 3 = 0\\
\end{array} \right.
\end{align}
and similarly for $\gamma_i$. Fig. \ref{ZeroModesC3} displays the quasiparticle tunneling process
described by these operators. For example, $\alpha_i$ describes the process in which a layer $2$ quasiparticle tunnels
around a twist defect and goes to layer $1$. The Klein factors are added so that these operators are zero modes:
\begin{align}
[\delta H_{tun}, \alpha_i ] = [\delta H_{tun},\beta_i] = [\delta H_{tun},\gamma_i] = 0.
\end{align}
Since $\phi_+ \equiv \phi_1 + \phi_2 + \phi_3$ is fixed everywhere,
there is a local constraint between the three zero modes similar to that in the $Z_2$ case (\it cf. \rm eq. (\ref{c2indZM}) ):
\begin{align}
\alpha_{k} \beta_{k} \gamma_{k} = e^{\frac{2\pi i}{m-l}} e^{-iQ_k},  \label{1dz3ChargeConstraint}
%&\alpha_{3k+1} \beta_{3k+1} \gamma_{3k+1} = e^{\frac{\pi i}{2(m-l)}} \nonumber \\
%&\alpha_{3k+2} \beta_{3k+2} \gamma_{3k+2} = e^{\frac{2\pi i}{m-l}} \nonumber \\
%&\alpha_{3k+3} \beta_{3k+3} \gamma_{3k+3} = e^{\frac{7\pi i}{2(m-l)}}
\end{align}
where $Q_k = \frac{1}{2\pi} (\phi_+(x_k) - \phi_+(x_{k-1}))$ is the charge on the $k$th defect,
where here $x_k$ refers to the region between the $(k-1)$th and $k$th defect.
In what follows, for simplicity we set $Q_k = 0$, as this does not affect
the topological properties of the defects.

Using the definition of the zero modes, we obtain the following algebra:
\begin{widetext}
\begin{align}
\alpha_n\alpha_{n+k}=\alpha_{n+k}\alpha_n e^{\frac{i2\pi}{m-l}}, & \;\; \beta_n\beta_{n+k}=\beta_{n+k}\beta_n e^{\frac{i2\pi}{m-l}}, \; \gamma_n\gamma_{n+k}=\gamma_{n+k}\gamma_n e^{\frac{i2\pi}{m-l}}, \nonumber \\
\alpha_n\beta_{n+k}=(-1)^{m-l}\beta_{n+k}\alpha_n e^{\frac{-i\pi}{m-l}}, & \;\; \beta_n\gamma_{n+k}=(-1)^{m-l}\gamma_{n+k}\beta_ne^{\frac{-i\pi}{m-l}},  \;\; \gamma_n\alpha_{n+k}=(-1)^{m-l}\gamma_{n+k}\alpha_n e^{\frac{-i\pi}{m-l}}, \nonumber \\
\alpha_n\gamma_{n+k}=(-1)^{m-l}\gamma_{n+k}\alpha_n e^{\frac{-i\pi}{m-l}}, &\;\; \beta_n\alpha_{n+k}=(-1)^{m-l}\beta_{n+k}\alpha_n e^{\frac{-i\pi}{m-l}}, \;\; \gamma_n\beta_{n+k}=(-1)^{m-l}\beta_{n+k}\gamma_n e^{\frac{-i\pi}{m-l}}, \nonumber \\
\alpha_n\beta_n=(-1)^{m-l}\beta_n\alpha_n e^{i\frac{\pi}{m-l}}, &\;\; \beta_n\gamma_n=(-1)^{m-l}\gamma_n\beta_n e^{i\frac{\pi}{m-l}}, \;\; \gamma_n\alpha_n=(-1)^{m-l}\alpha_n\gamma_n e^{i\frac{\pi}{m-l}}.
%\alpha_n\beta_{n+k}=-\beta_{n+k}\alpha_n e^{\frac{-i\pi}{m-l}}&, \; \alpha_n\gamma_{n+k}=-\gamma_{n+k}\alpha_n %e^{\frac{-i\pi}{m-l}} \nonumber \\
%\beta_n\alpha_{n+k}=-\beta_{n+k}\alpha_n e^{\frac{-i\pi}{m-l}}&, \; \beta_n\gamma_{n+k}=-\gamma_{n+k}\beta_n %e^{\frac{-i\pi}{m-l}} \nonumber \\
%\gamma_n\alpha_{n+k}=-\gamma_{n+k}\alpha_n e^{\frac{-i\pi}{m-l}}&, \; \gamma_n\beta_{n+k}=-\beta_{n+k}\gamma_n %e^{\frac{-i\pi}{m-l}} \nonumber \\
%\alpha_n\alpha_{n+k}=\alpha_{n+k}\alpha_n e^{\frac{i2\pi}{m-l}}&, \;  \alpha_n\beta_n=-\beta_n\alpha_n e^{\frac{i\pi}{m-l}}
%\nonumber \\
%\beta_n\beta_{n+k}=\beta_{n+k}\beta_n e^{\frac{i2\pi}{m-l}}&, \;  \beta_n\gamma_n=-\gamma_n\beta_n e^{\frac{i\pi}{m-l}}
%\nonumber \\
%\gamma_n\gamma_{n+k}=\gamma_{n+k}\gamma_n e^{\frac{i2\pi}{m-l}}&, \; \gamma_n\alpha_n=-\alpha_n\gamma_n e^{\frac{i\pi}{m-l}},
%\alpha_n\beta_{n+k}=-\beta_{n+k}\alpha_n e^{-i\pi /(m-l)}, \; \alpha_n\gamma_{n+k}=-\gamma_{n+k}\alpha_n e^{-i\pi/(m-l)} %\nonumber \\
%\beta_n\alpha_{n+k}=-\beta_{n+k}\alpha_n e^{-i\pi /(m-l)}, \; \beta_n\gamma_{n+k}=-\gamma_{n+k}\beta_n e^{-i\pi/(m-l)} %\nonumber \\
%\gamma_n\alpha_{n+k}=-\gamma_{n+k}\alpha_n e^{-i\pi /(m-l)}, \; \gamma_n\beta_{n+k}=-\beta_{n+k}\gamma_n e^{-i\pi/(m-l)} %\nonumber \\
%\alpha_n\alpha_{n+k}=\alpha_{n+k}\alpha_n e^{i2\pi/(m-l)}, \;  \alpha_n\beta_n=-\beta_n\alpha_n e^{i\pi/(m-l)}
%\nonumber \\
%\beta_n\beta_{n+k}=\beta_{n+k}\beta_n e^{i2\pi/(m-l)}, \;  \beta_n\gamma_n=-\gamma_n\beta_n e^{i\pi/(m-l)}
%\nonumber \\
%\gamma_n\gamma_{n+k}=\gamma_{n+k}\gamma_n e^{i2\pi/(m-l)}, \; \gamma_n\alpha_n=-\alpha_n\gamma_n e^{i\pi/(m-l)}
\end{align}
\end{widetext}
where $k>0$. The first line shows the algebra between the same zero mode operators located on different defects. The algebra of different operators on different sites is shown on the second and the third line, while in the last line, we write down the algebra between different operators on the same defect. The zero modes can be used to construct the quasiparticle Wilson loop operators:
%The irreducible representation of the quasiparticle Wilson loop algebra leads to the
%different topologically degenerate ground states. In the case of $n$ triplets of $Z_3$ defects, we find $3n$ copies of the magnetic algebra.
\begin{align}
W(a_{3k+1}) &= \alpha_{3k+1} \beta_{3k+2} \gamma_{3k+3},
\nonumber \\
W(b_{3k+1}) &= \gamma_{3k+3} \alpha_{3k+4} \beta_{3k+5},
\nonumber \\
W(a_{3k+2}) &= \beta_{3k+1} \gamma_{3k+2} \alpha_{3k+3},
\nonumber \\
W(b_{3k+2}) &= \alpha_{3+3} \beta_{3k+4} \gamma_{3k+5},
\nonumber \\
W(a_{3k+3}) &= -e^{\frac{i\pi}{m-l}} \alpha_{3k+2}^\dagger \alpha_{3k+3},
\nonumber \\
W(b_{3k+3}) &= -e^{\frac{i\pi}{m-l}}  \beta_{3k+2}^\dagger \beta_{3k+3},
\label{1dZ3WLdef}
\end{align}
where we have chosen the overall phases so that (see Appendix \ref{1dZ3WLidapp} )
\begin{align}
W(a_i)^{m-l}=W(b_i)^{m-l}=1.  \label{idz3wilson}
\end{align}
\begin{figure}
\centerline{
\includegraphics[width=3in]{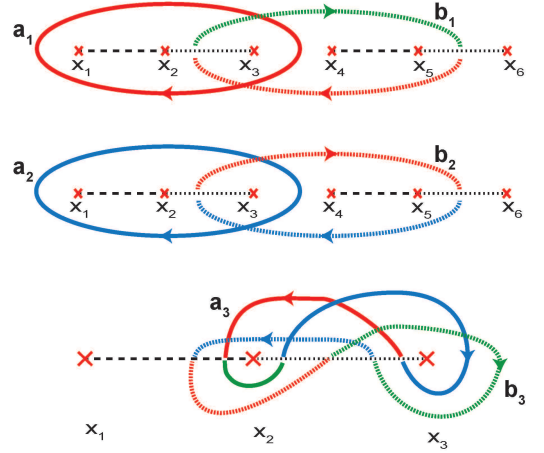}
}
\caption{Physical meaning of the quasiparticle Wilson loop operators.
\label{LoopsC3} }
\end{figure}
Fig. \ref{LoopsC3} displays the loops $a_i$ and $b_i$ along which the quasiparticle tunnels by the action of the Wilson loop operators $W(a_i)$ or $W(b_i)$.
These operators satisfy:
\begin{align}
\label{Z3alg}
W(a_{3k+i}) W(b_{3k+j}) = W(b_{3k+j}) W(a_{3k+i}) e^{\delta_{ij} 2\pi i/ (m-l)},
\end{align}
which leads to $|m-l|^{3n}$ states forming the irreducible representation of this algebra.
Adding an extra factor of $|m+2l|$ for the possible values of $e^{i (\phi_1 + \phi_2 + \phi_3)}$, we get a total of
$|(m+2l)(m - l)^{3n}|$ topologically degenerate states. This confirms again that the quantum dimension
of each $Z_3$ defect is $d_{Z_3} = |m-l|$.
\begin{figure*}
\centerline{
\includegraphics[width=6in]{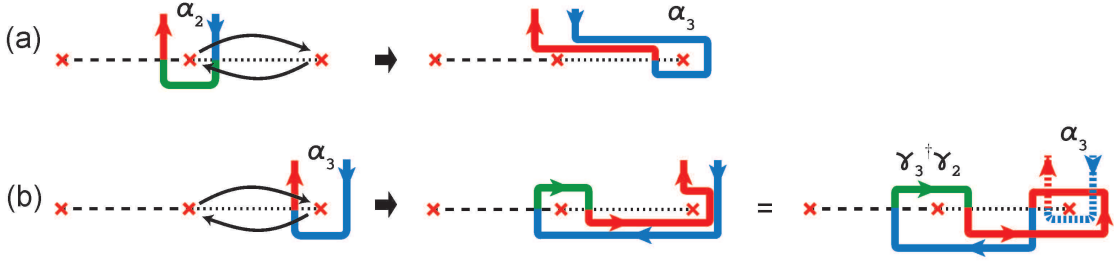}
}
\caption{Effect of a clockwise braid on the zero modes on the second and third dislocation. (a) $\alpha_2\rightarrow \alpha_3$ (b) $\alpha_3$ gets transformed into the combination of $\gamma_3^\dag\gamma_2$ and $\alpha_3$. Therefore, $\alpha_3\rightarrow\gamma_3^\dag\gamma_2\alpha_3$. Similar to the $Z_2$ case, the ordering ambiguity or additional phase factor can be fixed with additional constraints.
\label{1dZ3exchange} }
\end{figure*}

\begin{figure*}
\centerline{
\includegraphics[width=6in]{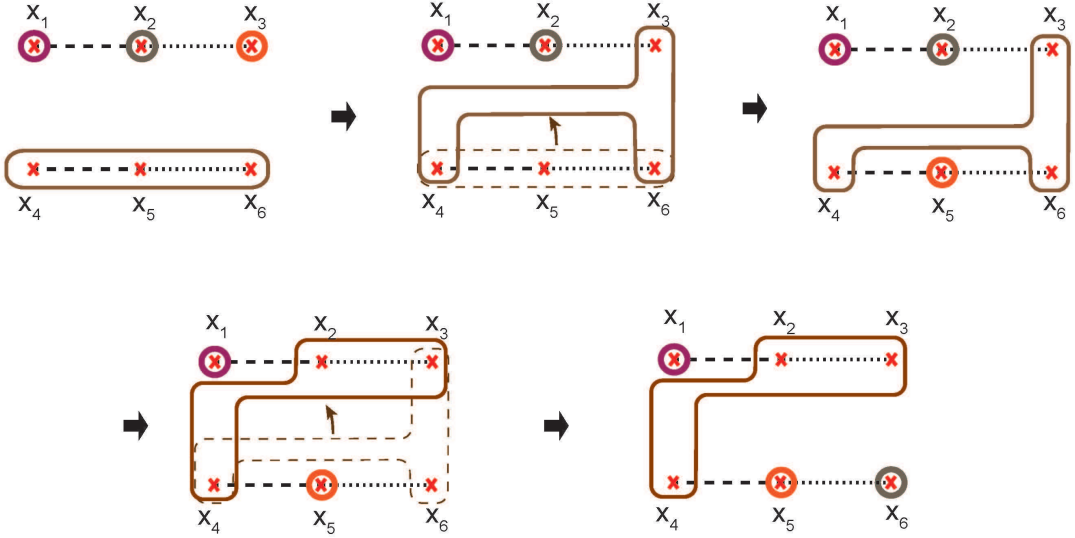}
}
\caption{Illustration of the 1D protocol for carrying out an effective braiding process of the $Z_3$ twist defects. The purple, orange and gray circles mark up the zero modes located on the corresponding defects.
\label{1dZ3braid} }
\end{figure*}

In order to calculate the braiding matrices, we need to understand how the zero mode
operators transform under a braiding process. From Fig. \ref{1dZ3exchange}, we see that
under a clockwise braiding of $2$ and $3$, $B_{23} \alpha_2 B_{23}^\dagger = e^{i\phi} \alpha_3$ and $B_{23} \alpha_3 B_{23}^\dagger =e^{i\theta} \gamma_3^\dagger \gamma_2  \alpha_3$, where $B_{23}$ is the braid matrix. The $Z_3$ layer cyclic symmetry implies that $B_{23} \beta_2 B_{23}^\dagger = e^{i\phi}\beta_3$, $B_{23}\beta_3 B_{23}^\dagger = e^{i\theta}\alpha_3^\dagger \alpha_2  \beta_3$, $B_{23} \gamma_2 B_{23}^\dagger=e^{i\phi} \gamma_3$ and $\gamma_3 B_{23}^\dagger =e^{i\theta} \beta_3^\dagger \beta_2  \gamma_3$. We can fix $e^{i\phi}=e^{-i\theta}$ by using the fact that the braiding operation should commute with the loop operator $W(a_1)$ and $W(a_2)$ that encircle these two defects. Using (\ref{1dz3ChargeConstraint}), we then have $e^{i3\phi}=1$. Thus, we find:
\begin{align}
B_{23} \alpha_2 B_{23}^\dagger &= e^{\frac{i2\pi k}{3}}\alpha_3, \;\; &B_{23} \alpha_3 B_{23}^\dagger = e^{\frac{-i2\pi k}{3}}\gamma_3^\dagger \gamma_2  \alpha_3,
\nonumber \\
B_{23} \beta_2 B_{23}^\dagger &= e^{\frac{i2\pi k}{3}}\beta_3, \;\; &B_{23} \beta_3 B_{23}^\dagger = e^{\frac{-i2\pi k}{3}}\alpha_3^\dagger \alpha_2  \beta_3,
\nonumber \\
B_{23} \gamma_2 B_{23}^\dagger &= e^{\frac{i2\pi k}{3}}\gamma_3, \;\; & B_{23} \gamma_3 B_{23}^\dagger = e^{\frac{-i2\pi k}{3}}\beta_3^\dagger \beta_2  \gamma_3,
\label{1dZ3braidOp}
\end{align}
where the integer $k=0,1,2$ is a phase ambiguity that is not yet fixed. Further, considering (\ref{idz3wilson}), we obtain $k=0$ when $m-l$ is not a multiple of 3. For $m-l$ being a multiple of 3, we still have 3 different choices of $k$, which is similar to the $Z_2$ case.

Now we choose a basis for the $|m-l|^{3n}$ different states, $|\{n_i\} \rangle$, for $n_i = 1, \cdots, |m-l|$ so that
\begin{align}
W(a_i)|\{n_j\} \rangle &= e^{2\pi i n_i/(m-l)} | \{n_j \} \rangle,
\nonumber \\
W(b_i)|\{n_j\} \rangle &= |\{ (n_j + \delta_{ij}) \% |m-l|\} \rangle.
\end{align}
Note that we can ignore the extra degeneracy of $|m+2l|$ associated with the possible eigenvalues of $e^{i\phi_+}$, as
they are independent of the braiding.
In this basis, we find that the braid matrix is:
\begin{align}
&\langle \{ n_i' \} | B_{23} | \{n_i\} \rangle \nonumber \\
&= \frac{ \delta_{n_1' n_1} \delta_{n_2' n_2}}{\sqrt{|m-l|}} e^{i\frac{4\pi k(n_3'-n_3)}{3}} e^{i \frac{\pi(n'^2_3 -2 n_3 n_3' -(m-l)n'_3)}{(m-l)} }.
\end{align}

The braid matrix $B_{23}$ can be viewed as an element of the mapping class group in the $U(1)_{m-l}$ CS
theory on a high genus surface. Observe that $B_{23}$ only has a non-trivial action on an $|m-l|$-dimensional subspace of the ground states,
associated with the states that form the $|m-l|$-dimesional irreducible representation of $W(a_3)$ and $W(b_3)$.
Now consider the Dehn twists $U_a$ and $U_b$, around the $a$ and $b$ cycles of the torus for $U(1)_{m-l}$ CS
theory (see (\ref{Z2Dehn1}), (\ref{Z2Dehn2}), (\ref{Z2Dehn2b})). When $m-l$ is odd, the action of $B_{23}$ in this relevant subspace is
\begin{align}
B_{23} = e^{i \theta} (U_a U_b)^\dagger,
\end{align}
where the phase $e^{i\theta}$ depends on details of the path. When $m-l$ is even,
\begin{align}
B_{23} = e^{i\theta} W(b_3)^{(m-l)/2}  (U_a U_b)^\dagger (W^\dagger(b_3))^{(m-l)/2}
\end{align}
Therefore, within this subspace, $B_{23}$ coincides with a sequence of Dehn twists in $U(1)_{m-l}$ CS.
When $B_{23}$ is not projected onto this subspace, it can be viewed as a sequence of Dehn twists
in $U(1)_{m-l}$ CS theory on a high genus surface. We will also see this result via the geometric
construction of the subsequent section.

Similar to the $Z_2$ case, we can also realize the braiding through a purely 1D protocol, without
continuously moving the defects in both dimensions. First, we define the following Hamiltonian:
\begin{align}
H_{abc}=-|t|(e^{\frac{-i2\pi\theta_{abc}}{m-l}}\alpha_a\beta_b\gamma_c+e^{\frac{-i2\pi\theta_{abc}}{m-l}} \beta_a\gamma_b\alpha_c
\nonumber \\
- e^{\frac{-i2\pi\theta_{abc}}{m-l}} e^{\frac{i\pi}{m-l}}\alpha_a^\dag \alpha_b+H.c.),
\label{1dZ3HamBlock}
\end{align}
which couples the zero modes located at the $a$, $b$ and $c$ defects. Here, we assume that $a<b<c$. In fact, all terms in this Hamiltonian commute with each other. The states in the subspace formed by the $a$, $b$ and $c$ defects are labeled by the discrete eigenvalues of the three operators $\alpha_a\beta_b\gamma_c$, $\beta_a\gamma_b\alpha_c$ and $\alpha_a^\dag \alpha_b$. So the ground state should take the eigenvalues that minimize each term (plus its Hermitian conjugate) in the Hamiltonian. For a generic $\theta_{abc}$, the degeneracy of Hilbert space associated with defect $a$, $b$ and $c$ is usually completely lifted. Although a finite set of values of $\theta_{abc}$ that will lead to accidental degeneracy in this Hilbert space exists, this set of values is carefully excluded in the following discussion.

Now, we can consider the following processes:
\begin{align}
H(\tau)=\left\{
\begin{array}{cc}
H_{3\rightarrow5}=(1-\tau)H_{456}+\tau H_{346},&\tau\in\left[0,1\right]  \\
H_{2\rightarrow6}=(2-\tau)H_{346}+(\tau-1) H_{234},&\tau\in\left[1,2\right] \\
\end{array}\right.
\label{1dZ3HamProtocol}
\end{align}
As is shown in Fig. \ref{1dZ3braid}, in the first half of the process $\tau\in[0,1]$, the Hamiltonian $H_{3\rightarrow5}$ takes the zero modes on defect $3$ to defect $5$. In the second half $\tau\in[1,2]$, the Hamiltonian $H_{2\rightarrow6}$ takes the zero modes on defect $2$ to $6$. We set $\theta_{456}=\theta_{234}$ so that up to a translation, this is a closed path in the Hamiltonian projected to the low energy subspace. This process effectively exchanges the zero modes at defects $2$ and $3$.

In order to understand the effect of these processes on the ground state subspace, we first observe that operators
\begin{align}
\mathcal{O}_1=\alpha_3\beta_4\gamma_5\alpha_6 & ,~\mathcal{O}'_1=\gamma_3\alpha_4\beta_5\gamma_6, \nonumber \\ %,~\mathcal{O}'_1=\beta_3\gamma_4\alpha_5\beta_6
\mathcal{O}_2=\alpha_2\beta_3\gamma_4\alpha_6 &, ~\mathcal{O}'_2=\gamma_2\alpha_3\beta_4\gamma_6 %~\mathcal{O}'_2=\beta_2\gamma_3\alpha_4\beta_6,
\end{align}
commute with the two processes, respectively:
\begin{align}
[H_{3\rightarrow5},\mathcal{O}_1]=[H_{3\rightarrow5},\mathcal{O}'_1]=0 \nonumber \\
[H_{2\rightarrow6},\mathcal{O}_2]=[H_{2\rightarrow6},\mathcal{O}'_2]=0
\end{align}
Using this, we can obtain the effect of these processes on the zero modes. We let $\mathcal{P}_{a\rightarrow b}(\tau)$ be the projector onto the ground state sector of $H_{a\rightarrow b}$. First, we define integers $k_1$ and $k_2$ such that
\begin{align}
&\theta_{456}\in(k_1-1/2,k_1+1/2),\nonumber \\
&\theta_{346}\in(k_2-1/2,k_2+1/2),
\end{align}
where $\theta_{456},\theta_{346}\neq\mathbb{Z}+1/2$ is assumed to avoid accidental degeneracy in the Hilbert space associated with these defects.
Then, following the same logic as the 1D protocol for the $Z_2$ twist defect, we can use the projection $\mathcal{P}_{3\rightarrow5}(\tau)\mathcal{O}_1\mathcal{P}_{3\rightarrow5}(\tau)$ and $\mathcal{P}_{3\rightarrow5}(\tau)\mathcal{O}'_1\mathcal{P}_{3\rightarrow5}(\tau)$ to study the evolution of zero modes in the first half of the process ($\tau$ from $0$ to $1$) and consider $\mathcal{P}_{3\rightarrow5}(\tau)\mathcal{O}_2\mathcal{P}_{3\rightarrow5}(\tau)$ and $\mathcal{P}_{3\rightarrow5}(\tau)\mathcal{O}'_2\mathcal{P}_{3\rightarrow5}(\tau)$ for the second half ($\tau$ from $1$ to $2$). We find that
\begin{align}
&\mathcal{P}_{3\rightarrow5}(0)\mathcal{O}_1\mathcal{P}_{3\rightarrow5}(0)=e^{\frac{i2\pi k_1}{m-l}}\alpha_3, \nonumber \\
%&\mathcal{P}_{3\rightarrow5}[0]\mathcal{O}'_1\mathcal{P}_{3\rightarrow5}[0]=e^{\frac{-i4\pi k_1}{m-l}}\beta_3, \nonumber \\
&\mathcal{P}_{3\rightarrow5}(0)\mathcal{O}'_1\mathcal{P}_{3\rightarrow5}(0)=e^{\frac{i2\pi k_1}{m-l}}\gamma_3, \nonumber \\
&\mathcal{P}_{3\rightarrow5}(1)\mathcal{O}_1\mathcal{P}_{3\rightarrow5}(1)=e^{\frac{i2\pi
k_2}{m-l}}\gamma_6^\dag\gamma_5\alpha_6, \nonumber \\
%&\mathcal{P}_{3\rightarrow5}[1]\mathcal{O}'_1\mathcal{P}_{3\rightarrow5}[1]=e^{\frac{i2\pi
%k_2}{m-l}}\alpha_6^\dag\alpha_5\beta_6, \nonumber \\
&\mathcal{P}_{3\rightarrow5}(1)\mathcal{O}'_1\mathcal{P}_{3\rightarrow5}(1)=e^{\frac{-i4\pi k_2}{m-l}} \beta^\dag_6\beta_5 \gamma_6,
\end{align}
where we have used the relations $(\alpha_3\beta_4\gamma_6)^{m-l}=(\beta_3\gamma_4\alpha_6)^{m-l}=1$ and $\alpha_i\beta_i\gamma_i=e^{\frac{2\pi i}{m-l}}$ (assuming the charge on each dislocation is $0$). Also, we have
\begin{align}
&\mathcal{P}_{2\rightarrow6}(1)\mathcal{O}_2\mathcal{P}_{2\rightarrow6}(1)=e^{\frac{i2\pi k_2}{m-l}}\alpha_2, \nonumber \\
%&\mathcal{P}_{2\rightarrow6}[0]\mathcal{O}'_2\mathcal{P}_{2\rightarrow6}[0]=e^{\frac{-i4\pi k_2}{m-l}}\beta_2, \nonumber \\
&\mathcal{P}_{2\rightarrow6}(1)\mathcal{O}'_2\mathcal{P}_{2\rightarrow6}(1)=e^{\frac{i2\pi k_2}{m-l}}\gamma_2, \nonumber \\
&\mathcal{P}_{2\rightarrow6}(2)\mathcal{O}_2\mathcal{P}_{2\rightarrow6}(2)=e^{\frac{i2\pi k_1}{m-l}}\alpha_6, \nonumber \\
%&\mathcal{P}_{2\rightarrow6}[1]\mathcal{O}_2\mathcal{P}_{2\rightarrow6}[1]=e^{\frac{i2\pi k_1}{m-l}}\beta_6, \nonumber \\
&\mathcal{P}_{2\rightarrow6}(2)\mathcal{O}'_2\mathcal{P}_{2\rightarrow6}(2)=e^{\frac{-i4\pi k_1}{m-l}}\gamma_6.
\end{align}
Notice that, in the second half of the process, $\alpha_6$ and $\gamma_6$ commute with the Hamiltonian and thus remain unchanged. Similar to the $Z_2$ case, after the process is over, we can relabel defect 5 as defect 2 and defect 6 as defect 3 . Therefore, with the relations $\alpha_i\beta_i\gamma_i=e^{\frac{2\pi i}{m-l}}$, we can write down these transformations of the zero modes in terms of the braid matrix:
\begin{align}
&B_{23} \alpha_2 B_{23}^\dagger = e^{\frac{i2\pi (k_1-k2)}{m-l}}\alpha_3,  \nonumber \\
&B_{23} \alpha_3 B_{23}^\dagger = e^{\frac{i2\pi (k_2-k_1)}{m-l}}\gamma_3^\dagger \gamma_2  \alpha_3, \nonumber \\
&B_{23} \beta_2 B_{23}^\dagger = e^{\frac{i2\pi (k_1+2k_2)}{m-l}}\beta_3, \nonumber \\
&B_{23} \beta_3 B_{23}^\dagger = e^{\frac{i2\pi (2k_1+k2)}{m-l}}\alpha_3^\dagger \alpha_2  \beta_3, \nonumber \\
&B_{23} \gamma_2 B_{23}^\dagger = e^{\frac{-i2\pi (2k_1+k_2)}{m-l}}\gamma_3, \nonumber \\
&B_{23} \gamma_3 B_{23}^\dagger = e^{\frac{-i2\pi (k_1+2k_2)}{m-l}}\beta_3^\dagger \beta_2  \gamma_3.
\end{align}
This result is slightly different from (\ref{1dZ3braidOp}). The reason is that when we restrict the discussion to
the same charge sector, in which the relations $\alpha_i\beta_i\gamma_i=e^{\frac{2\pi i}{m-l}}$ hold for any $i$ (see (\ref{1dz3ChargeConstraint})) ,
the Hamiltonian (\ref{1dZ3HamProtocol}) of this 1D protocol breaks the $Z_3$ cyclic layer symmetry for generic
$k_1$ and $k_2$. If $m-l$ is not a multiple of 3, only when $k_1=k_2=0$ is the $Z_3$ layer symmetry
restored. If $m-l$ is a multiple of 3, the condition of the symmetry is $k_1=0$ and $k_2=0,\pm (m-l)/3$.
Once the symmetry is restored, this result agrees with (\ref{1dZ3braidOp}). On the other hand, if we
choose to loosen the charge constraints but insist on the $Z_3$ layer symmetry of (\ref{1dZ3HamProtocol})
(by adding other terms related to the original terms by cyclic permutation into the definition of
(\ref{1dZ3HamBlock})), then the charge on each dislocation changes generically during the "braiding"
process dictated by the 1D protocol Hamiltonian (\ref{1dZ3HamProtocol}).

\subsection{Bulk geometric picture}
\label{Z3geoSec}

\begin{figure*}
\centerline{
\includegraphics[width=6in]{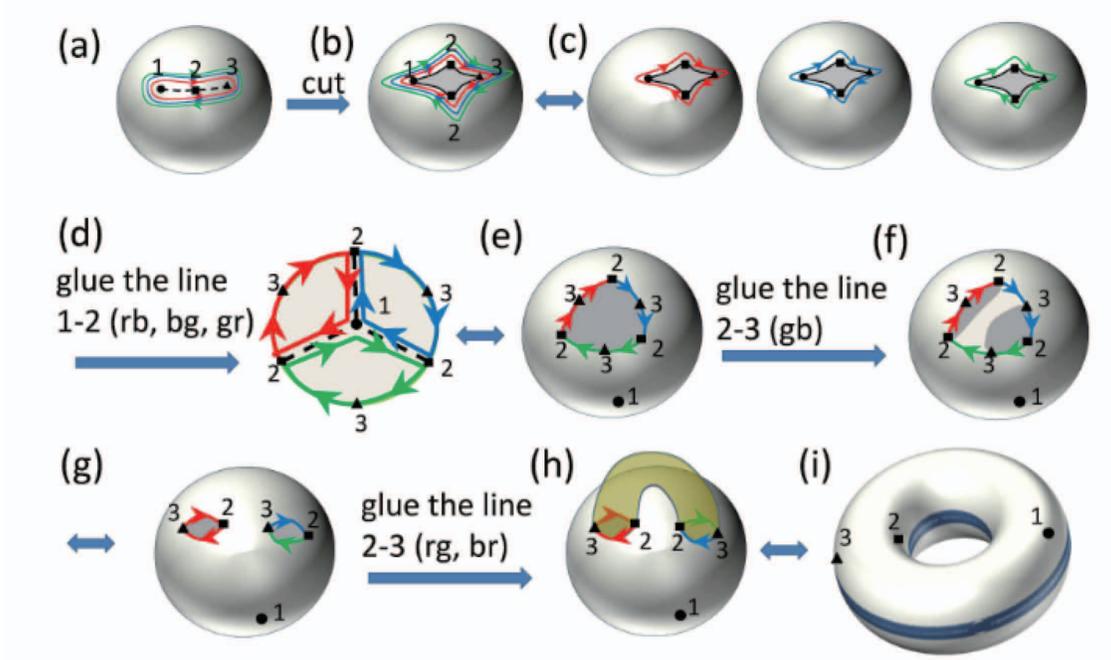}
}
\caption{Illustration of the equivalence between a sphere with three $Z_3$ twist defects and a torus. (a) and (b) shows the sphere with twist defects cut along the branch-cut line connecting defects $1,2,3$. The red, blue and green lines denote the edge states of the three layers along the cut. After the cutting, the system in (b) is equivalent to three decoupled systems of the three layers shown in (c). Then the three layers are glued along the branch-cut line after a twist. Along line $1-2$ the three layers are glued in the pattern of red-to-blue (rb), blue-to-green (bg) and green-to-red (gr), in which the first and second colors are the layer index of the lower and upper edges respectively. The result of this gluing is shown in (d) which is equivalent to (e). Then the green and blue layers are glued along $2-3$ line, shown in panels (f) and (g). Finally, the rest of the edge states in (g) are glued following the rule of red-to-green, green-to-blue, which leads to the torus shown in (h) and (i).
\label{geoZ3} }
\end{figure*}

\begin{figure}[b]
\centerline{
\includegraphics[width=3.5in]{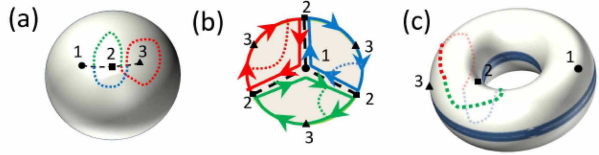}
}
\caption{A loop around two $Z_3$ twist defects $2$ and $3$ (the dotted lines) on the sphere (a) is topologically equivalent to a nontrivial loop on the torus shown in (c). This is the same loop that appears in Fig. \ref{1dZ3exchange} (b). The color of the dotted line indicates the layer indices. This equivalence is obtained by following the steps shown in Fig. \ref{geoZ3}. Panel (b) shows an intermediate steps of the topological deformation which corresponds to Fig. \ref{geoZ3} (d). Other intermediate steps are omitted.
\label{geoZ3braid} }
\end{figure}

%XLinsert: Pls copy this section to the latest version.

When $l=0$ in the $K$ matrix in Eq. (\ref{KZ3}), the three layers are decoupled and it is possible to
understand the $Z_3$ twist defects by directly mapping the tri-layer system with twist defects to a
high genus surface, similar to the bilayer $Z_2$ case. The mapping is probably too complicated for
practical purposes, and we will only discuss it in the simplest case of a sphere with three $Z_3$
defects, as an illustration of the general situation (as in the $Z_2$ case, we expect the mapping to
also be possible when $l\neq 0$, though this is even more complicated). Illustration of the mapping is given in
Fig. \ref{geoZ3}. We consider a cut-and-glue procedure to describe the defects. Cutting
the sphere along a line connecting the defects $1,2,3$ leads to chiral edge states of the
three layers around the cut, as is shown in Fig. \ref{geoZ3} (a) where the three layers are
labeled by red, blue and green lines. As is illustrated in Fig. \ref{geoZ3} (b), the system with
the cut is equivalent to three decoupled layers, each of which is topologically a disk. Then the
twist defects are created by gluing different layers shifted by a $Z_3$ operation. Along the
branch-cut between defects $1$ and $2$ the left-movers on the lower edge in the red (r),
blue (b) and green (g) layers are glued with the right movers on the upper edge in the b,g,r layers,
respectively. Along the branch-cut between $2$ and $3$, the layers are glued in the opposite
fashion so that r,b,g left-movers are glued to g,r,b right-movers, respectively. This gluing
can be done in three steps. First, the layers are glued along the branch-cut between $1$
and $2$, which glues the three disks into one disk, as is shown in Fig. \ref{geoZ3} (d) and
(e). Secondly, along the branch-cut between $2$ and $3$ the green left-mover was glued
to the blue right-mover, which leads to a sphere with two punctures shown in
Fig. \ref{geoZ3} (g). Thirdly, the remaining red and blue left-moving edge states are glued
to green and red right-movers, respectively, which glues the punctured sphere to a torus.

We can also see from the geometric picture that the Wilson loop operators indeed correspond
to a non-trivial loops on the torus. For example, the loop corresponding to the operator
$\gamma_2^\dag \gamma_3$ is shown in Fig. \ref{geoZ3braid} (a). In the mapping this loop is
mapped to a nontrivial loop on the torus, as is illustrated in Fig. \ref{geoZ3braid} (b) and (c).
The braiding of the defect $2$ and $3$ will deform this loop to another non-trivial loop on
the torus. Thus, we know that the braiding of the defects correspond to a non-trivial element
of the mapping class group.

\section{Deconfining the genons: orbifold states}
\label{orbifoldSec}

So far, we have considered twist defects (genons), labeled by a symmetry group element $g \in G$, where $G$
is a symmetry of the topological quantum numbers of a topological phase. These genons are
\it extrinsic \rm defects of the system, and separating them costs energy that grows with the distance
between them, similar to vortices in a superfluid. It is possible to deconfine these defects to turn them
into intrinsic quasiparticles of a neighboring topological phase. One way to do this is by finding a
way to gauge the symmetry $G$. Such a gauging process corresponds to orbifolding in the edge CFT.\cite{dijkgraaf1989}
%Such a construction in CFT, where a discrete symmetry is gauged, is called an orbifold CFT.
In this section we will discuss the relation between the braiding statistics of the genons in the confined (ungauged) and
deconfined (gauged) states. The difference between confined and deconfined genons shown below will also further clarify 
the intrinsic difference between projective and linear representations of the braid group.

In the case of $Z_2$ twist defects studied in Section \ref{Z2twocomp}, one can gauge the $Z_2$ symmetry associated with exchanging
the two $U(1)$ gauge fields. Physically, we expect that this can be done either by proliferating double
twist defects,\footnote{This expectation is based on previously known examples where $Z_2$ gauge theory emerges by proliferating
double vortices in a superfluid. } or by condensing anyons\cite{barkeshli2010prl}. This leads to the $U(1) \times U(1) \rtimes Z_2$ CS theory studied in Ref. \onlinecite{barkeshli2010}, which describe a set of novel orbifold FQH states\cite{barkeshli2011orb} and ``twisted'' $Z_n$ topological
states.\cite{barkeshli2010twist} There, the $Z_2$ twist defects simply become finite-energy $Z_2$ vortices, which carry
quantum dimension $\sqrt{|m-l|}$. These states contain an Abelian quasiparticle that carries the $Z_2$ gauge charge;
when it condenses, the system undergoes a $Z_2$ Higgs transition in the 3D Ising universality class, and
we obtain the Abelian $U(1) \times U(1)$ theories.

There is a close relation between the braiding of the $Z_2$ vortices in the orbifold states and the braiding
of the extrinsic defects in the Abelian states. To understand this, let us briefly review the properties of the orbifold
states. We recall that in the $U(1) \times U(1)$ CS theory, for $n$ pairs of $Z_2$ twist defects on a sphere,
the ground state degeneracy is $|m-l|^{n-1}$. When the $Z_2$ symmetry is gauged, $\alpha_n$ of these states are
$Z_2$ invariant, while $\beta_n$ are $Z_2$ non-invariant, where $\alpha_n + \beta_n = |m-l|^{n-1}$, and it was
found\cite{barkeshli2010} that
\begin{align}
\alpha_n = \left\{ \begin{array}{ccc}
(|m-l|^{n-1} + 2^{n-1})/2 & \text{ for } & |m-l| \text{ even } \\
(|m-l|^{n-1} + 1)/2 & \text{ for } & |m - l | \text{ odd }
\end{array} \right.
\end{align}

\begin{align}
\beta_n = \left\{ \begin{array}{ccc}
(|m-l|^{n-1} - 2^{n-1})/2 & \text{ for } & |m-l| \text{ even } \\
(|m-l|^{n-1} - 1)/2 & \text{ for } & |m - l | \text{ odd }
\end{array} \right.
\end{align}

Therefore, in the orbifold/twisted states, where we gauge the $Z_2$, the number of ground states in the presence
of $n$ pairs of $Z_2$ vortices is $\alpha_n$. This result allows us to deduce the fusion rules\cite{barkeshli2010, barkeshli2011orb}
\begin{align}
(\gamma \times \bar{\gamma})^{n} = \alpha_n \mathbb{I} + \beta_n \phi + \cdots,
\end{align}
where $\phi$ is the Abelian quasiparticle that carries the $Z_2$ gauge charge,
$\gamma$ represents the $Z_2$ vortex, and $\bar{\gamma}$ its conjugate. Therefore,
the braid matrix for the $Z_2$ vortices in this case will have dimension $\alpha_n$
instead of $|m-l|^{n-1}$. \it Thus for $| m -l | > 2$, the braid matrices for genons in the confined and deconfined
states are not the same, and even have a different dimension. \rm

As we found in Section \ref{Z2twocomp}, the braiding of the genons corresponds to Dehn twists of the $U(1)_{m-l}$ CS theory. Under the
action of the Dehn twists, $Z_2$ invariant states are mapped to $Z_2$ invariant states, and similarly for
the $Z_2$ non-invariant states. Therefore, the $|m-l|^{n-1}$ dimensional braid matrix of the genons,
when restricted to the $Z_2$ invariant subspace, will be equivalent to the $\alpha_n$-dimensional braid
matrix of the $Z_2$ vortices.

%A more striking example of the difference between braiding of the twist defects in the confined vs. deconfined states
%occurs in the case of $Z_2$ twist defects in the Ising $\times$ Ising theory. In Sec. \ref{Z2na}, we found that
%the $Z_2$ twist defects in the Ising $\times$ Ising state have quantum dimension $d_{Z_2} = 2$, and their braiding
%can be used for universal TQC. When the $Z_2$ symmetry of the Ising $\times$ Ising theory is gauged,
%we obtain a new topological state, whose topological properties can be deduced by studying the
%$[Ising \times Ising]/Z_2$ orbifold CFT. The $[Ising \times Ising]/Z_2$ orbifold CFT has central charge
%$c = 1$, and since all $c = 1$ rational CFTs are classified,\cite{ginsparg1988,kiritsis1989} we know that the total quantum dimension squared
%$D^2$ is integer. According to the conjecture of \Ref{rowell2009}, it follows that this state cannot be universal for TQC.
%Therefore, if the conjecture of \Ref{rowell2009} is true, the braiding of the $Z_2$ twist defects is sufficiently different after
%the $Z_2$ is gauged, that the state is no longer universal for TQC.

In the single-component case, described by $U(1)_N$ CS theory, we considered $Z_2$ twist
defects associated with $a \rightarrow -a$, which give defects with quantum dimension $\sqrt{2N}$.
We can consider gauging the $Z_2$ symmetry in this theory, to obtain a CS theory with disconnected
gauge group $U(1) \rtimes Z_2 = O(2)$.\cite{moore1989, barkeshli2010} The edge CFT of such a CS theory is the $U(1)_N/Z_2$ orbifold
theory.\cite{moore1989, dijkgraaf1989} When $N$ is even, the $Z_2$ vortices in this theory have quantum dimension $\sqrt{N/2}$,
while when $N$ is odd, they have quantum dimension $\sqrt{N}$. This provides an example where gauging
the symmetry associated with the twist defect appears to also change its quantum dimension!

Let us briefly explain the reason for this change in quantum dimension. When the genons are confined, and have
quantum dimension $\sqrt{2N}$, part of this quantum dimension in this context is actually protected by a
particle number symmetry. When $N$ is odd (\it ie \rm for fermionic states), the $\sqrt{2}$ factor is protected by the fermion parity
symmetry. When $N$ is even, then we can think of the quantum dimension as $\sqrt{2N} = 2 \sqrt{N/2}$; the factor of
$2$ in this quantum dimension can be shown to be protected by a boson parity number symmetry. Notice that
in the edge theory (see Sec \ref{Z2onecomp}), only boson pairing terms are added, so boson parity is still a symmetry of the problem. When these symmetries
are completely broken (by e.g. coupling the system to an external particle reservoir), then we see that the fermionic case gives a quantum dimension $\sqrt{N}$, while the bosonic
case gives a quantum dimension $\sqrt{N/2}$, which coincides with the result obtained by considering
the $O(2)$ CS theory, discussed above, where the genons are deconfined.

In the $Z_3$ case, gauging the $Z_3$ symmetry leads to $[U(1) \times U(1) \times U(1)] \rtimes Z_3$
CS theory. We expect the relation between the $Z_3$ vortices in that theory and the $Z_3$ twist
defects in the $U(1) \times U(1) \times U(1)$ theory to be similar to the case of the two-component
Abelian states.

\section{Discussion}
\label{discSec}

In this paper, we have studied the possibility of extrinsic twist defects, associated with a symmetry of the
topological quantum numbers of a topologically ordered state. We have explicitly studied several different
examples, including various kinds of $Z_2$ and $Z_3$ twist defects in Abelian and non-Abelian topological states.
In all of the cases we have studied, we found that the projective braiding statistics of the twist defects are related to adiabatic
modular transformations of a topological state on a high genus surface, and we have developed several different
ways of explicitly computing the braiding of the twist defects. The close and precise relation of the twist defects to
topological states on a high genus surface leads us to refer to them as genons.

From a mathematical point of view, the mapping we found from braiding operations of $Z_2$ twist defects to Dehn twists defines a group homomorphism between the braid group $B_{2n}$ and the mapping class group of genus $n-1$ surfaces. (More generally, for each $m\geq 2$, a different homomorphism is defined from the braid group to the mapping class group by the braiding of $Z_m$ twist defects.) The relation between braid groups and mapping class groups has been studied extensively in the mathematics literature (see {\it e.g.} Ref. \cite{birman1969,margalit2009}), but as far as we can understand, the homomorphism we discuss seems to be different from the existing ones in the literature.

%It should be noted that the braid group $B_n$ is known to be isomorphic to the mapping class group of a $n$-punctured disk (see {\it e.g.} {\bf ref}), which is different from the homomorphism we discuss. It is interesting to ask whether these two types of mappings are related to each other.

Most importantly, the braiding statistics of the genons are \it projective\rm, meaning that they are only well-defined up to
an overall phase. The reason for this is that the genons are not intrinsic dynamical quasiparticles of the system.
This fact allows the possibility of fundamentally novel behavior, because the projective non-abelian braiding statistics are not subject
to the same stringent constraints as true non-abelian quasiparticles. In this paper, we have found several examples of this.
The simplest case of $Z_2$ twist defects in Abelian states have quantum dimension $\sqrt{n}$ for integer $n$. For $n = 2$, these
defects are Majorana fermions and their braiding is the same as the braiding of Ising anyons, up to an overall phase. However, as we discussed in
Sec. \ref{orbifoldSec}, the braiding of the confined defects for $n > 2$ is inequivalent to the braiding that is obtained when the genons are deconfined,
because the dimension of the braid matrices are different in the two cases. In some cases, we even found that after the symmetry is
gauged in the theory so that the genons become deconfined, it appears that their quantum dimension can also change.

While the non-abelian genons in Abelian states have been argued to be insufficient for universal TQC,\cite{lindner2012} we
have found that $Z_2$ genons introduced into non-abelian states can be used for universal TQC, even when the host
non-Abelian state is non-universal for TQC.
%These defects have integer quantum dimension, while true non-abelian quasiparticles
%are believed to allow for universal TQC if and only if their quantum dimension is not the square root of an integer. This again highlights
%the fundamental difference in possibilities between projective non-abelian statistics and true non-abelian statistics.

In this work, the twist defects are point defects in the sense that far away from the defects, no local operation can distinguish the presence
of the defect, because the defects are associated with symmetries of the topological order. It is possible to consider a more general class
of defects, which are different kinds of domain walls between different topological states. Then, junctions between different kinds of
domain walls may induce some non-trivial topological degeneracy and cannot be viewed as point defects at all.
We leave the detailed study of such more general possibilities for future work.

\vskip.2in
\it Acknowledgements \rm We would like to thank Michael Freedman, Zhenghan Wang, and Xiao-Gang Wen
for helpful discussions, and the Institute for Advanced Study at Tsinghua University for hospitality
while part of this work was completed. This work was supported by the Simons Foundation (MB) and the David and Lucile Packard Foundation (CMJ and XLQ).

\appendix

\appendix
\section{Ground state degeneracy with $Z_3$ twist defects via minima of tunneling operators}
\label{1dZ3GSDapp}
Similar to the $Z_2$ case, we shall obtain the ground state degeneracy of the system by considering the degenerate minima of the tunneling operators in (\ref{1dZ3tunnel}). Since the Klein factors are defined differently with $m-l$ even and odd, we will separate the discussion into two pieces. We assume $g<0$ throughout the discussion.

\subsubsection{$m-l$ even}

In the absence of twist defects, the whole 1+1D edge is effectively in the region $A$. The operators $\mathcal{K}_1$, $\mathcal{K}_2$ and $\mathcal{K}_3$, which have eigenvalues $\pm 1$, commute with each other. We will project the Hilbert space onto the sector where $\mathcal{K}_1$, $\mathcal{K}_2$ and $\mathcal{K}_3$ take specific values, since states with different values of $\mathcal{K}_{1,2,3}$ are not topologically degenerate. For our convenience, we choose the sector where
$\mathcal{K}_{1,2,3}=1$. Then the conditions of minimizing $\delta H_{tun}$ in (\ref{1dZ3tunnel}) are
\begin{align}
m\phi_1+l \phi_2 + l \phi_3= 2\pi n_1^A, \nonumber \\
l\phi_1+m \phi_2 + l \phi_3= 2\pi n_2^A,  \nonumber \\
l\phi_1+l \phi_2 + m \phi_3= 2\pi n_3^A,
\label{1dZ3minimaCond1}
\end{align}
where $n_{a,b,c}\in Z$. These equations can be put into a diagonal form:
\begin{align}
&(m+2l)\phi_+ = 2\pi (n_1^A+n_2^A+n_3^A) \equiv 2\pi p_+ , \nonumber \\
&(m-l)\phi_{(1,-2,1)}= 2\pi (n_1^A- 2 n_2^A +n_3^A) \equiv 2\pi p_{(1,-2,1)}, \nonumber \\
&(m-l)\phi_{(1,0,-1)}= 2\pi (n_1^A-n_3^A) \equiv 2\pi p_{(1,0,-1)}.
\label{1dZ3minimaCond2}
\end{align}
Different values that $\phi_I$'s are pinned to represent different degenerate ground states. Since $\phi_I$'s are subject to the periodicity $\phi_I\thicksim\phi_I+2\pi$, we should actually use $e^{i \phi_+}$, $e^{i \phi_{(1,-2,1)}}$ and $e^{i \phi_{(1,0,-1)}}$ to label the ground states, as shown in Sec. \ref{Z3CFT}. From (\ref{1dZ3minimaCond2}), we obtain (\ref{1dZ3DegLabel}):
\begin{align}
e^{i \phi_+} |\vec{p} \rangle &= e^{\frac{2\pi i p_+}{m+2l}} |\vec{p} \rangle,
\nonumber \\
e^{i \phi_{(1,-2,1)}} |\vec{p}\rangle &= e^{\frac{i 2\pi p_{(1,-2,1)}}{m - l }} |\vec{p} \rangle,
\nonumber \\
e^{i \phi_{(1,0,-1)}} |\vec{p}\rangle &= e^{\frac{i 2\pi p_{(1,0,-1)}}{m - l }} |\vec{p} \rangle.
\nonumber
\end{align}
We find that $e^{i \phi_+}$ can take $m+2l$ different values, while each of $e^{i \phi_{(1,-2,1)}}$ and $e^{i \phi_{(1,0,-1)}}$ has $m-l$ different choices. Thus, in the absence of twist defects, the ground state degeneracy of the system on a torus is $|(m+2l)(m-l)^2|$, as expected from the $U(1)\times U(1)\times U(1)$ CS theory.

Now we consider the case with $n$ triplets of twist defects. First of all, by the same reasoning above, we can project the Hilbert space onto the sector where $\mathcal{K}_I$, $\mathcal{K}'_I$ and $\mathcal{K}''_I$ take specific values, since they commute with each other. We choose the sector where $\mathcal{K}_I=\mathcal{K}'_I=\mathcal{K}''_I=1$. The ground states are labeled by the maximally commuting set of gauge invariant operators:
\begin{align}
\{ & e^{i \phi_+(x_{A_i})},~e^{i \phi_{(1,-2,1)}(x_{A_i})},~e^{i \phi_{(1,0,-1)}(x_{A_i})},~e^{i \phi_+(x_{B_i})},  \nonumber \\
&  e^{i \phi_+(x_{C_i})},~e^{i \phi'_{(1,0,-1)}(x_{B_i})+i \phi''_{(-1,1,0)}(x_{C_i})} \}.
\label{1dZ3MaxCommut}
\end{align}
For all operators in region $A$, the analysis above follows straightforwardly leading to the conclusion that each of $e^{i \phi_+(x_{A_i})}$ can be pinned to one of $m+2l$ different values, while each of $e^{i \phi_{(1,0,-1)}(x_{A_i})}$ and $e^{i \phi_{(1,-2,1)}(x_{A_i})}$ can be pinned to one of $|m-l|$ values. To determine the possible values of $e^{i \phi_+(x_{B_i})}$, $e^{i \phi_+(x_{C_i})}$ and $e^{i \phi'_{(1,0,-1)}(x_{B_i})+i \phi''_{(-1,1,0)}(x_{C_i})}$, we need to write down the conditions minimizing $\delta H_{tun}$ in (\ref{1dZ3tunnel}) in region $B$ and $C$:
\begin{align}
m\phi'_1+l \phi'_2 + l \phi'_3= 2\pi n_1^B, \nonumber \\
l\phi'_1+m \phi'_2 + l \phi'_3= 2\pi n_2^B, \nonumber \\
l\phi'_1+l \phi'_2 + m \phi'_3= 2\pi n_3^B, \nonumber \\
m\phi''_1+l \phi''_2 + l \phi''_3= 2\pi n_1^C, \nonumber \\
l\phi''_1+m \phi''_2 + l \phi''_3= 2\pi n_2^C, \nonumber \\
l\phi''_1+l \phi''_2 + m \phi''_3= 2\pi n_3^C,
\label{1dz3minimaCond3}
\end{align}
which can be rewritten in the diagonal form (including the position index)
\begin{align}
&(m+2l)\phi_+(x_{B_i}) = 2\pi (n_1^{B_i}+n_2^{B_i}+n_3^{B_i})  , \nonumber \\
&(m-l)\phi'_{(1,-2,1)}(x_{B_i})= 2\pi (n_1^{B_i}- 2 n_2^{B_i} +n_3^{B_i}) , \nonumber \\
&(m-l)\phi'_{(1,0,-1)}(x_{B_i})= 2\pi (n_1^{B_i}-n_3^{B_i}) , \nonumber \\
&(m+2l)\phi_+(x_{C_i}) = 2\pi (n_1^{C_i}+n_2^{C_i}+n_3^{C_i})  , \nonumber \\
&(m-l)\phi'_{(1,-2,1)}(x_{C_i})= 2\pi (n_1^{C_i}- 2 n_2^{C_i} +n_3^{C_i}) , \nonumber \\
&(m-l)\phi'_{(1,0,-1)}(x_{C_i})= 2\pi (n_1^{C_i}-n_3^{C_i}) .
\end{align}
Here, $n_{1,2,3}^{B_i},n_{1,2,3}^{C_i}\in Z$. Thus, we find each of $e^{i \phi_+(x_{B_i})}$ and $e^{i \phi_+(x_{C_i})}$ can
be pinned to one of $m+2l$ different values, while $e^{i \phi'_{(1,0,-1)}(x_{B_i})+i \phi''_{(-1,1,0)}(x_{C_i})}$ is pinned
to one of $|m-l|$ values. This counting gives a total of $|(m+2l)(m-l)|^{3n}$ states. Considering the charge
constraint on each defect, which manifests itself in the difference between $\phi_+$ in
different regions (see Sec. \ref{Z3CFT}), the topological ground state degeneracy is $|(m+2l)(m-l)^{3n}|$.

\subsubsection{$m - l$ odd}
In the absence of the twist defects, the analysis of ground state degeneracy is similar to the case above. Since the operators $\mathcal{K}_{1,2,3}$ commute with each other, we can restrict the system to be in the sector where $\mathcal{K}_{1,2,3}=1$. Thus, the condition for minimizing $\delta H_{tun}$ is the same as (\ref{1dZ3minimaCond1}). In the same fashion, we use the operators $e^{i \phi_+}$, $e^{i \phi_{(1,-2,1)}}$ and $e^{i \phi_{(1,0,-1)}}$ to distinguish the degenerate ground states. The ground state degeneracy on torus without twist defects is, therefore, given by the same formula $|(m+2l)(m-l)^2|$.

Now we consider the case with $n$ triplets of twist defects. The treatment will be slightly more subtle than the $m-l$ even case. We still make the projection $\mathcal{K}_{1,2,3}=1$ and identify the maximally commuting set of operators as (\ref{1dZ3MaxCommut}). Thus, all the analysis within region $A$ stays the same. Each of $e^{i \phi_+(x_{A_i})}$ can be pinned to one of $m+2l$ different values, while each of $e^{i \phi_{(1,0,-1)}(x_{A_i})}$ and $e^{i \phi_{(1,-2,1)}(x_{A_i})}$ is pinned to one of $|m-l|$ values. However, in region $B$ and $C$, $\mathcal{K}'_{1,2,3}$ and $\mathcal{K}''_{1,2,3}$ cannot be fixed because they anti-commute with $\mathcal{K}_{1,2,3}$. With different $\mathcal{K}'_I$ and $\mathcal{K}''_I$ values, the cosine potentials in (\ref{1dZ3tunnel}) will pin the $\phi$ fields to different phases which can be written in a compact form:
\begin{align}
m\phi'_1(x_{B_i})+l \phi'_2(x_{B_i}) + l \phi'_3(x_{B_i})= 2\pi n_1^{B_i}+\frac{1-\mathcal{K}'_1}{2}\pi, \nonumber \\
l\phi'_1(x_{B_i})+m \phi'_2(x_{B_i}) + l \phi'_3(x_{B_i})= 2\pi n_2^{B_i}+\frac{1-\mathcal{K}'_2}{2}\pi, \nonumber \\
l\phi'_1(x_{B_i})+l \phi'_2(x_{B_i}) + m \phi'_3(x_{B_i})= 2\pi n_1^{B_i}+\frac{1-\mathcal{K}'_3}{2}\pi, \nonumber \\
m\phi''_1(x_{C_i})+l \phi''_2(x_{C_i}) + l \phi''_3(x_{C_i})= 2\pi n_1^{C_i}+\frac{1-\mathcal{K}''_1}{2}\pi, \nonumber \\
l\phi''_1(x_{C_i})+m \phi''_2(x_{C_i}) + l \phi''_3(x_{C_i})= 2\pi n_2^{C_i}+\frac{1-\mathcal{K}''_2}{2}\pi, \nonumber \\
l\phi''_1(x_{C_i})+l \phi''_2(x_{C_i}) + m \phi''_3(x_{C_i})= 2\pi n_3^{C_i}+\frac{1-\mathcal{K}''_3}{2}\pi.
\label{1dZ3minimaCond1odd}
\end{align}
The sum of the first three equations in (\ref{1dZ3minimaCond1odd}) yields
\begin{align}
&(m+2l)\phi_+(x_{B_i}) \nonumber \\
&=2\pi (n_1^{B_i}+n_2^{B_i}+n_3^{B_i}) +\frac{3-\mathcal{K}'_1-\mathcal{K}'_2-\mathcal{K}'_3}{2} \pi.
\end{align}
Since, by definition, $\mathcal{K}'_1\mathcal{K}'_2\mathcal{K}'_3=\mathcal{K}_1\mathcal{K}_2\mathcal{K}_3$, $\frac{3-\mathcal{K}'_1-\mathcal{K}'_2-\mathcal{K}'_3}{2} \pi$ must be a multiple of $2\pi$ in the sector $\mathcal{K}_{1,2,3}=1$. Hence, we can package the right hand side of this equation above into a single integer $p_+^{B_i}$ times $2\pi$. And we can conclude that each of $e^{i \phi_+(x_{B_i})}$ is pinned to one of $m+2l$ different values. The same argument applies for $e^{i \phi_+(x_{C_i})}$. From (\ref{1dZ3minimaCond1odd}), we can also have
\begin{align}
&(m-l)(\phi'_{(1,0,-1)}(x_{B_i})+ \phi''_{(-1,1,0)}(x_{C_i})) \nonumber \\
&=2\pi(n_1^{B_i}-n_3^{B_i}-n_1^{C_i}+n_2^{C_i})+\frac{4-\mathcal{K}'_1+\mathcal{K}'_3+\mathcal{K}''_1-\mathcal{K}''_2}{2} \pi.
\end{align}
Due to the fact that $\mathcal{K}'_1\mathcal{K}''_2=-\mathcal{K}_1\mathcal{K}_2$ and $\mathcal{K}'_3\mathcal{K}''_1=-\mathcal{K}_1\mathcal{K}_3$, we have $\mathcal{K}'_1+\mathcal{K}''_2=0$ and $\mathcal{K}'_3+\mathcal{K}''_1=0$ in the sector where $\mathcal{K}_{1,2,3}=1$. So we can rewrite the equation as
\begin{align}
(m-l)(\phi'_{(1,0,-1)}(x_{B_i})+ \phi''_{(-1,1,0)}(x_{C_i})) \nonumber \\
=2\pi(n_1^{B_i}-n_3^{B_i}-n_1^{C_i}+n_2^{C_i}+1).
\end{align}
Therefore, $e^{i \phi'_{(1,0,-1)}(x_{B_i})+i \phi''_{(-1,1,0)}(x_{C_i})}$ is pinned to one of $|m-l|$ different values, just like that in the case with $m-l$ even. Then, this counting will give us in total of $|(m+2l)(m-l)|^{3n}$ states. Considering the charge constraints on each defect, we conclude the topological ground state degeneracy is $|(m+2l)(m-l)^{3n}|$.

\section{Identities of Wilson loop operators}
\label{1dZ3WLidapp}
The definition of the Wilson loop operators are given in (\ref{1dZ3WLdef}). Now we will prove (\ref{idz3wilson}). To simplify notation, we consider the case with $k=0$ in (\ref{1dZ3WLdef}). The Wilson loop operators are written as
\begin{align}
W(a_{1}) = \alpha_{1} \beta_{2} \gamma_{3},~&
W(b_{1}) = \gamma_{3} \alpha_{4} \beta_{5},
\nonumber \\
W(a_{2}) = \beta_{1} \gamma_{2} \alpha_{3},~&
W(b_{2}) = \alpha_{3} \beta_{4} \gamma_{5},
\nonumber \\
W(a_{3}) = -e^{\frac{i\pi}{m-l}} \alpha_{2}^\dagger \alpha_{3},~&
W(b_{3}) = -e^{\frac{i\pi}{m-l}}  \beta_{2}^\dagger \beta_{3}.
\end{align}
Expressed in terms of boson fields, we rewrite $W(a_{1})$ as
\begin{align}
W(a_{1})=e^{i\phi_1(x_{A_1})}e^{-i\phi_1(x_{A_2})}
\end{align}
With the assumption that the charge on each dislocation is $0$, $\phi_+$ is pinned to the one single value globally. From (\ref{1dZ3minimaCond1}), ,we can obtain
\begin{align}
(m-l) \phi_1(x_{A_1})-(m-l) \phi_1(x_{A_2})=2\pi (n_1^{A_1}-n_1^{A_2}).
\end{align}
Thus,
\begin{align}
W(a_{1})^{m-l}=e^{i (m-l) \phi_1(x_{A_1})}e^{-i (m-l) \phi_1(x_{A_2})}=1.
\end{align}
The proof for $W(b_{1})^{m-l}=W(a_{2})^{m-l}=W(b_{2})^{m-l}=1$ is parallel. Now, we will focus the discussion on $W(a_{3})$. First, we write
\begin{align}
\alpha_2^\dag\alpha_3 &=e^{i \phi''_3(x_{C_1})}e^{-i \phi'_3(x_{B_1})} e^{i \phi''_2(x_{C_1})} e^{-i \phi_2(x_{A_2})}  \nonumber \\
& =e^{-i\frac{\pi}{m-l}} e^{-i \phi'_3(x_{B_1})} e^{i \phi''_2(x_{C_1})+i \phi''_3(x_{C_1})} e^{-i \phi_2(x_{A_2})}
\nonumber \\
& =e^{-i\frac{\pi}{m-l}} e^{-i \phi'_3(x_{B_1})} e^{-i \phi''_1(x_{C_1})} e^{-i \phi_2(x_{A_2})} e^{i \phi_+}
\end{align}
Notice that the 4 operators in this expression commute with each other. Thus,
\begin{align}
(\alpha_2^\dag\alpha_3)^{m-l}= & (-1) e^{-i (m-l)\phi'_3(x_{B_1})} e^{-i (m-l) \phi''_1(x_{C_1})} \nonumber \\
& \times e^{-i (m-l) \phi_2(x_{A_2})} e^{i (m-l) \phi_+}
\end{align}
Rewriting the second formula in (\ref{1dZ3minimaCond1}) as $(m-l)\phi_2+l\phi_+=2\pi n_2^A$ , we have
\begin{align}
e^{-i (m-l) \phi_2(x_{A_2})}=e^{il\phi_+}.
\end{align}
For the operator $e^{-i (m-l)\phi'_3(x_{B_1})} e^{-i (m-l) \phi''_1(x_{C_1})}$, the even/odd-ness of $m-l$ matters. For $m-l$ even, we obtain from (\ref{1dz3minimaCond3}) that $(m-l)(\phi'_3(x_{B_1})+\phi''_1(x_{C_1}))+2l \phi_+=2\pi Z$, where $Z$ means an integer, which leads to
\begin{align}
e^{-i (m-l)\phi'_3(x_{B_1})} e^{-i (m-l) \phi''_1(x_{C_1})}=e^{i 2l \phi_+}
\end{align}
For $m-l$ odd,
\begin{align}
(m-l)(\phi'_3(x_{B_1})+\phi''_1(x_{C_1}))+2l \phi_+=2\pi Z+\pi-\frac{\mathcal{K}'_3+\mathcal{K}''_1}{2}.
\end{align}
Since $\mathcal{K}'_3\mathcal{K}''_1=-\mathcal{K}_3\mathcal{K}_1=-1$, we have $\mathcal{K}'_3+\mathcal{K}''_1=0$. So, in this case,
\begin{align}
e^{-i (m-l)\phi'_3(x_{B_1})} e^{-i (m-l) \phi''_1(x_{C_1})}=-e^{i 2l \phi_+}
\end{align}
Thus, for general $m-l$,
\begin{align}
e^{-i (m-l)\phi'_3(x_{B_1})} e^{-i (m-l) \phi''_1(x_{C_1})}=(-1)^{m-l}e^{i 2l \phi_+}.
\end{align}
Now, we have
\begin{align}
(\alpha_2^\dag\alpha_3)^{m-l}=(-1)^{m-l+1}e^{i (m+2l) \phi_+}=(-1)^{m-l-1}.
\end{align}
Considering the phase factor in the definition of $W(a_3)$, we have proven
that $W(a_3)^{m-l}=1$. And, by cyclic symmetry of the layer index, we also have $W(b_3)^{m-l}=1$.
Besides the the Wilson Loop operators defined in (\ref{1dZ3WLdef}), we will derive similar identities for the Wilson Loop operators $\alpha_3\beta_4\gamma_6$ and $\beta_3\gamma_4\alpha_6$ which is useful in the 1D protocol. Written in terms of the boson fields,
\begin{align}
\alpha_3\beta_4\gamma_6 & =e^{i\phi''_2(x_{C_1})}e^{-i\phi'_2(x_{B_2})}e^{i\phi''_1(x_{C_2})}e^{-i\phi_1(x_{A_3})} \nonumber \\ &=e^{i\frac{\pi}{m-l}}e^{-i\phi'_2(x_{B_2})}e^{i\phi''_2(x_{C_1})+i\phi''_1(x_{C_2})}e^{-i\phi_1(x_{A_3})}.
\end{align}
Thus, by taking the $m-l$th power of this expression and rearranging the operators while taking all possible commutator in account, we have
\begin{align}
&(\alpha_3\beta_4\gamma_6)^{m-l} \nonumber \\
&=(-1)^{m-l}e^{-i(m-l)\phi'_2(x_{B_2})}e^{i(m-l)(\phi''_2(x_{C_1})+\phi''_1(x_{C_2}))} \nonumber \\
&~~\times e^{-i(m-l)\phi_1(x_{A_3})}, \nonumber \\
&=(-1)^{m-l}e^{-i(m-l)\phi'_2(x_{B_2})}e^{i(m-l)(\phi''_2(x_{C_2})+\phi''_1(x_{C_2}))} \nonumber \\
&~~\times e^{-i(m-l)\phi_1(x_{A_3})} e^{i(m-l)(\phi''_2(x_{C_1})-\phi''_2(x_{C_2}))}, \nonumber \\
&=(-1)^{m-l}e^{-i(m-l)\phi'_2(x_{B_2})}e^{-i(m-l)\phi''_3(x_{C_2})} \nonumber \\
&~~\times e^{-i(m-l)\phi_1(x_{A_3})} e^{i(m-l)(\phi''_2(x_{C_1})-\phi''_2(x_{C_2}))} e^{i(m-l)\phi_+}. \nonumber \\
\end{align}
In this derivation, we have used the fact that $e^{i(m-l)(\phi''_2(x_{C_1})-\phi''_2(x_{C_2}))}$ commute with $e^{-i(m-l)\phi'_2(x_{B_2})}$, $e^{-i(m-l)\phi_1(x_{A_3})}$ and $e^{i(m-l)\phi_+}$. From (\ref{1dZ3minimaCond1}) or (\ref{1dZ3minimaCond1odd}), we have $e^{i(m-l)(\phi''_2(x_{C_1})-\phi''_2(x_{C_2}))}=1$. And following a parallel discussion
of $W(a_3)$, we obtain $e^{-i(m-l)\phi'_2(x_{B_2})}e^{-i(m-l)\phi''_3(x_{C_2})}=(-1)^{m-l}e^{i 2l \phi_+}$ and $e^{-i(m-l)\phi_1(x_{A_3})}=e^{i l \phi_+}$. Therefore, we conclude
%\begin{align}
$(\alpha_3\beta_4\gamma_6)^{m-l}=e^{i (m+2l) \phi_+}=1$.
%\end{align}
By cyclic symmetry of layers, we also obtain $(\beta_3\gamma_4\alpha_6)^{m-l}=1$.

%\bibliography{TI}

\end{document}